# Are eHMIs always helpful? Investigating how eHMIs interfere with pedestrian behavior on multi-lane streets: An eye-tracking virtual reality experiment

Yun Ye[1,2,3], Zexuan Li[1,2], Panagiotis Angeloudis[3], S.C. Wong[4], Jian Sun[5], Haoyang Liang[5†]

[1] Faculty of Maritime and Transportation, Ningbo University, Ningbo, China

[2] Collaborative Innovation Center of Modern Urban Traffic Technologies, Southeast University, Nanjing, China

[3] Centre for Transport Engineering and Modelling, Department of Civil and Environmental Engineering, Imperial College London, London, UK

[4] Department of Civil Engineering, The University of Hong Kong, Hong Kong, China

[5] College of Transportation Engineering and the Key Laboratory of Road and Traffic Engineering, Ministry of Education, Tongji University, Shanghai, China

† **Correspondence to**: lianghy@connect.hku.hk

**Acknowledgement:** This research was supported by the Zhejiang Provincial Natural Science Foundation of China (Grant No. LQN25E080011), Ningbo Natural Science Foundation (Grant No. 2024J440), National "111" Centre on Safety and Intelligent Operation of Sea Bridges (Project No. D21013), and Research Grants Council of the Hong Kong Special Administrative Region, China (Project No. T32-707/22-N). The fifth author was also supported by the Francis S Y Bong Professorship in Engineering. The experiment was facilitated by the TransCAVE Lab at Tongji University (http://202.120.189.178:5021/#/serviceSystem, Account/PW: guest/tjutju01). The funders had no role in the study design, data collection and processing, manuscript preparation, or decision to publish.

**Authorship contribution statement**

**Yun Ye:** Conceptualization, Formal analysis, Investigation, Methodology, Funding acquisition, Supervision, Writing-original draft. **Zexuan Li:** Formal analysis, Software, Data curation, Investigation, Methodology, Visualization, Writing-original draft. **Panagiotis Angeloudis:** Validation, Writing-review & editing. **S.C. Wong:** Writing-review & editing, Funding acquisition. **Jian Sun:** Writing-review & editing, Resources. **Haoyang Liang:** Conceptualization, Project administration, Resources, Writing-review & editing.

**Declaration of competing interest**

The authors declare that they have no known competing financial interests or personal relationships that could have appeared to influence the work reported in this paper.

# Are eHMIs always helpful? Investigating how eHMIs interfere with pedestrian behavior on multi-lane streets: An eye-tracking virtual reality experiment

**ABSTRACT**

Appropriate communication is crucial for efficient and safe interactions between pedestrians and autonomous vehicles (AVs). External human–machine interfaces (eHMIs) on AVs, which can be categorized as allocentric (displaying vehicle motion–related information) or egocentric (guiding pedestrian behavior), are considered a promising solution. While the effectiveness of eHMIs has been extensively studied, in complex environments, such as unsignalized multi-lane streets, their potential to interfere with pedestrian crossing behavior remains underexplored. Hence, a virtual reality–based experiment was conducted to examine how different types of eHMIs displayed on AVs affect the crossing behavior of pedestrians in multi-lane streets environments, with a focus on the gaze patterns of pedestrians during crossing. The results revealed that the presence of eHMIs significantly influenced the cognitive load on pedestrians and increased the possibility of distraction, even misleading pedestrians in cases involving multiple AVs on multi-lane streets. Notably, allocentric eHMIs induced higher cognitive loads and greater distraction in pedestrians than egocentric eHMIs. This was primarily evidenced by longer gaze time and higher proportions of attention for the eHMI on the interacting vehicle, as well as a broader distribution of gaze toward vehicles in the non-interacting lane. However, misleading behavior was mainly triggered by eHMI signals from yielding vehicles in the non-interacting lane. Under such asymmetric signal configurations, egocentric eHMIs resulted in a higher misjudgment rate than allocentric eHMIs. These findings highlight the importance of enhancing eHMI designs to balance the clarity and consistency of the displayed information across different perspectives, especially in complex multi-lane traffic scenarios. This study provides valuable insights regarding the application and standardization of future eHMI systems for AVs.

# 1. Introduction

While walking remains a fundamental mode of daily travel for many individuals, pedestrians have consistently been vulnerable participants in road traffic systems. According to the Global Status Report on Road Safety 2023 (WHO, 2023), pedestrians account for 21% of all traffic fatalities globally, ranking third among all road user groups. Among the primary factors contributing to pedestrian injuries and fatalities, distraction or impaired perception is particularly significant (Wang et al., 2020). This underscores the criticality of studying pedestrian behavior as a key element in enhancing the safety of vehicle–pedestrian interactions.

In traditional road systems, vehicle–pedestrian interactions typically rely on explicit communication methods, such as traffic signals and crosswalks, to allocate right-of-way and determine safe crossing opportunities. However, in areas without traffic lights or crosswalks, implicit forms of communication, such as an understanding of vehicle speed and distance, eye contact, and physical gestures, become the primary means for pedestrians and vehicles to negotiate right-of-way. In recent years, the rapid advancement of autonomous driving technology has increased the presence of autonomous vehicles (AVs) across road systems, rendering implicit communication more challenging when traffic lights and crosswalks are absent (Guo et al., 2022). According to AV classification standards, Level 4/5 AVs require no driver responsibility during dynamic monitoring or driving tasks, potentially reducing the risks associated with interactions between human drivers and pedestrians (Betz et al., 2022). However, the significantly different design of AVs may conflict with the longstanding habits of pedestrians developed from interacting with conventional vehicles, potentially influencing their road-crossing decisions.

Installing external human–machine interfaces (eHMIs), which transmit vehicle state information and behavioral cues, on AVs has become essential for AV–pedestrian communication. Various eHMI designs have been proposed, including anthropomorphic displays, facial expressions, text, light bars, and projections (Chang et al., 2018). Existing eHMIs can be divided into two categories based on whether the information is conveyed from the vehicle's or pedestrian's perspective: egocentric eHMIs and allocentric eHMIs (Fig. 1). Egocentric eHMIs provide advisory information from the pedestrian's perspective, such as "Walk" or "Stop", while allocentric eHMIs offer referential advice from the vehicle's perspective, such as "Driving" or "Braking". As a novel form of vehicle–pedestrian interaction, eHMIs demonstrate the potential to improve traffic efficiency and enhance the sense of safety for pedestrians crossing the road. However, most studies thus far have focused on interactions between pedestrians and single AVs (Tran et al., 2021). When multiple AVs are equipped with eHMIs, pedestrians may receive conflicting information (Song et al., 2023). Such situations may complicate crossing decisions for pedestrians by hindering their ability to interpret each vehicle's intentions, potentially even misguiding their crossing behavior.

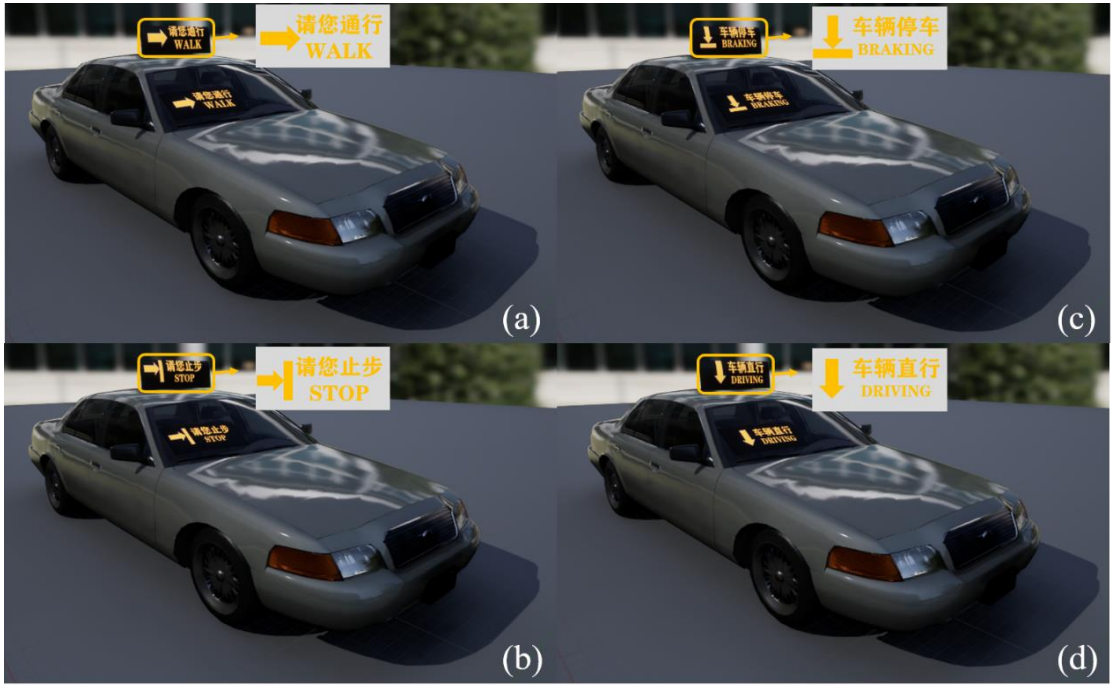

**Fig. 1.** Types of eHMIs based on communication perspective: Egocentric eHMIs: (a) Walk; (b) Stop; Allocentric eHMIs: (c) Braking; (d) Driving.

Accordingly, the interactions of pedestrians with egocentric and allocentric eHMIs in multi-vehicle contexts were investigated to assess how eHMIs influence cognitive load, cause distractions, and mislead pedestrians, and how these factors affect the crossing risk for pedestrians. This was achieved by developing an interactive virtual reality (VR) platform equipped with eye-tracking devices to capture the gaze patterns of pedestrians. The experiments included three forms of communication for pedestrians—no eHMI, allocentric eHMI, and egocentric eHMI—and two AV interaction strategies—i.e., yielding and non-yielding; this framework enabled a comprehensive analysis of the key factors influencing pedestrian cognitive load, distraction, and misguidance in multi-vehicle scenarios. Specifically, this study aimed to investigate the following issues:

- Pedestrian cognitive load in a multi-lane environment with multiple eHMIs displayed on AVs;
- The distraction caused by information from eHMIs on AVs in adjacent lanes at pedestrian crossings;
- Misleading communication in multi-lane environments with multiple eHMIs on separate AVs;
- Pedestrian crossing risk in a multi-lane environment with multiple eHMIs on separate AVs.

The remainder of this paper is organized as follows: Section 2 reviews existing eHMI applications in human–vehicle interactions. Section 3 outlines the experimental design and data analysis methods used in this study. Section 4 presents the results. Section 5 discusses the findings. Finally, Section 6 concludes the paper.

## 2. Literature review

### 2.1 Studies on eHMIs

With the rapid advancement of autonomous driving technology, eHMIs have garnered growing attention considering their potential to assist pedestrians in identifying the states and intentions of vehicles. To date, researchers have proposed various eHMI designs, including light signals, text displays, projections, and anthropomorphic features (Chang et al., 2017; Clamann et al., 2017; Mührmann et al., 2019; Ferenchak et al., 2022). These diverse systems aim to compensate for the absence of non-verbal communication cues traditionally present in interactions between pedestrians and human-driven vehicles. Barendse et al. (2019) categorized eHMIs based on two distinct "perspectives": egocentric communication and allocentric communication. Egocentric eHMIs provide pedestrians with intuitive guidance focused on their own actions, such as advising them to "Walk" or "Stop", whereas allocentric communication informs pedestrians about the current behavior or status of the vehicle, such as by displaying signals such as "Braking" or "Driving." This distinction not only reflects differences in eHMI design priorities but also has significant implications for how pedestrians perceive environmental risks, allocate attention, and make crossing decisions. However, research on the combined presentation of these two information perspectives remains relatively limited.

A large body of empirical research has confirmed that eHMIs can be effective in enhancing pedestrian safety and boosting decision-making confidence. Studies widely agree that eHMIs help improve the situational awareness of pedestrians, reduce their uncertainty during the decision-making process, and increase their efficiency in crossing the street (Izquierdo et al., 2023). Furthermore, several studies have identified eHMIs as a promising solution to bridge the communication gap between pedestrians and AVs, particularly in right-of-way negotiations, where they can enhance the trust of pedestrians in AVs (Habibovic et al., 2018; Kooijman et al., 2019; Dey et al., 2021). However, eHMIs do not always produce positive outcomes in all traffic scenarios. Some studies have highlighted potential safety concerns when eHMIs are used in complex traffic environments. For instance, Eisma et al. (2019) noted that strong visual stimuli, such as projections or dynamic light strips, may distract pedestrians and reduce their awareness of the surrounding traffic context. Additionally, overly prescriptive eHMI messages may prompt pedestrians to cross prematurely, even in unsafe situations (Kaleefathullah et al., 2022; Lee et al., 2024). Furthermore, when eHMI messages are misunderstood by pedestrians or when vehicle kinematic cues alone suffice for safe decision-making, the presence of eHMIs may lead to cognitive confusion and heightened environmental uncertainty (Weber et al., 2019; de Winter et al., 2022).

**2.2 Interference by eHMIs with pedestrian crossing behaviors**

At signalized intersections, AV–pedestrian interactions are clearly structured. However, in unsignalized or priority-lacking areas, communication between pedestrians and AVs becomes ambiguous (Rasouli et al., 2019). Such environments expose pedestrians to interference that impairs attention, risk perception, and decision-making. Most studies have focused on the distraction and cognitive load perceived by pedestrians during street crossing. However, in complex traffic environments, misleading behaviors, such as pedestrians misjudging safe crossing opportunities due to ambiguous or conflicting eHMI signals, resulting from these factors also form an integral part of the interference mechanism. Research on misleading behaviors remains relatively limited, especially in multi-agent interaction scenarios.

Distraction effects, one of the most common research topics, primarily consist of behaviors that involve shifting attention during street crossing (Bungum et al., 2005). Attention-diverting factors during crossing significantly raise accident risk (Mohammed, 2021). To better understand pedestrian behavior and the associated risks in complex environments, researchers have extensively explored how these distraction factors affect pedestrian responses and decisions in an autonomous driving context. In addition to the active behaviors of pedestrians, the visual complexity of the surrounding environment is one of the key causes of unconscious misallocation of attention resources, especially in tasks that rely heavily on visual input (Rosenholtz et al., 2007). Schwebel et al. (2024) categorized pedestrian distraction into "technical" (e.g., devices) and "social" (e.g., conversation). Social distractions (e.g., group interactions) may be more dangerous, as they cause pedestrians to miss key traffic cues. Hossain et al. (2024) found that all distraction types increase both accident risk and injury severity. Distracted pedestrians show unsteady gait, slow reactions, and jaywalking (Wang et al., 2022; Campisi et al., 2024; Raoniar & Maurya, 2024). Observations show that distracted pedestrians cross slower, with less visual attention and lower safety (O'Dell et al., 2023; Krishna et al., 2024).

Cognitive load is another important factor in pedestrian interference. In traffic behavior research, "cognitive load" refers to the mental burden of processing information that an individual experiences while navigating road systems (Dommes, 2019). Pedestrians rely heavily on cognition to process visuals, assess risk, and decide quickly in complex settings (Stavrinos et al., 2018). Excessive information or task complexity increases load, impairing attention, comprehension, and decisions. The advent of VR technology has enabled researchers to systematically examine pedestrian behavior under controlled experimental conditions (Ye et al., 2020; Ye et al., 2023). VR interference tasks reveal how complex settings affect crossing behavior and cognitive allocation (Schneider, 2019; Tapiro et al., 2020; Tian et al., 2022). Weiss et al. (2022) manipulated cognitive load levels within a dual-task experimental paradigm and found that even under high-load conditions, the ability of pedestrians to understand eHMI communication remained relatively unaffected. However, they also highlighted that multi-source interference in real-world environments could still undermine the effectiveness of eHMIs.

While existing research has extensively verified the heightened risks associated with pedestrian distraction, misleading information represents another critical, yet often overlooked, factor that threatens pedestrian safety. Cognitive and social psychology define misleading information as stimuli that cause misinterpretation due to ambiguity or deception (Pillai and Fazio, 2021), weakening pedestrians' risk perception and judgment (Kaleefathullah, 2022). In AV–pedestrian interactions, unclear or conflicting eHMI signals can mislead pedestrians (Holländer et al., 2019). Few studies examine how eHMIs can both distract and mislead pedestrians. In complex, multi-vehicle traffic environments, poorly designed eHMIs may not only divert pedestrian attention but also distort their interpretation of right-of-way negotiations or vehicle yielding intentions, thus exacerbating the risk of accidents.

Table 1 summarizes the literature related to the effects of interference on pedestrian crossing behaviors. In brief, while eHMIs show considerable potential in optimizing AV–pedestrian interactions and improving pedestrian safety, existing research has largely focused on specific designs or isolated scenarios. Whether the concurrent presence of different eHMI perspectives may inadvertently distract or mislead pedestrians in multi-lane, multi-agent environments remains to be systematically explored.

**Table 1.** Summary of studies on factors interfering with pedestrian crossing.

| Category | | Major findings | Authors |
|---|---|---|---|
| Cognitive load | VR cognitive interference task | High visual clutter increases missed crossing opportunities for adults and children | Tapiro et al. (2020) |
| | | Auditory–cognitive increases the probability of unsafe decisions; time pressure leads to smaller acceptable gaps and riskier crossing attempts | Tian et al. (2022) |
| | Dual-task experiment | An increase in cognitive load has no significant effect on how fast or accurately pedestrians understand eHMI icons | Weiss et al. (2022) |
| Distraction effect | Attention distraction | Distracted pedestrians may need more crossing time, which affects their safety | Mohammed (2021) |
| | Types of distraction | The occurrence of different distraction types is influenced by multiple factors and demonstrates strong contextual dependence | Schwebel et al. (2024) |

| | | Different types of pedestrian distractions are context-dependent and significantly linked to crash severity and location | Hossain et al. (2024) |
|---|---|---|---|
| | | Tasks leading to visual and auditory distraction significantly impair pedestrian response and alter brain activation patterns, increasing safety risks | Wang et al. (2022) |
| | | Distracted pedestrians walk slower and disregard signs | Campisi et al. (2024) |
| | Effects of distraction | Distracted pedestrians are more likely to violate signals, walk slower, and experience near-misses | Raoniar & Maurya (2024) |
| | | Distraction increases the probability of unsafe interactions, while higher walking speeds increase risk; visual behavior insights are crucial for safety interventions | Krishna et al. (2024) |
| | | Distracted pedestrians show less safe behaviors, look less at traffic, and take longer to cross | O'Dell et al. (2023) |
| Misleading effect | Potential misleading effects of eHMIs | Repeated exposure to misleading information increases belief in falsehoods | Holländer et al. (2019) |

### 2.3 Aim of this study

This study aimed to investigate pedestrian crossing behavior and the underlying risk mechanisms when pedestrians interact with multiple eHMIs displayed on AVs on multi-lane streets, focusing on four key dimensions: cognitive load, distraction effects, misleading behavior, and crossing risk. Specifically, the research explored how different types of eHMI information influence the attention allocation, risk perception, and decision-making behavior of pedestrians through these interference mechanisms in complex, multi-agent traffic environments. Data were collected through subjective questionnaires and VR experiments, which involved capturing the eye movements, crossing behavior, and self-reported feedback of pedestrians during their interactions with AVs featuring both egocentric and allocentric eHMIs on multi-lane streets. A generalized linear mixed model (GLMM) was employed to analyze the influence of multidimensional factors on the psychological states, behavioral responses, and crossing risks of the pedestrians. The findings reveal the potential interference and safety implications of eHMIs in complex AV–pedestrian interaction scenarios and provide both theoretical insights and practical guidance for optimizing eHMI design and informing real-world traffic policy development.

### 3. Methodology

The experimental design encompasses both VR scenario development and questionnaire formulation. Objective behavioral and eye-tracking data were collected from pedestrians through the VR experiment, while subjective cognitive data were obtained via questionnaires. The GLMM was then applied to both the objective behavioral and subjective cognitive data to assess how the eHMIs interfered with pedestrian behavior in terms of cognitive load, distraction effect, misleading effect, and crossing risk in a multi-lane street environment. Fig. 2 illustrates the overall framework.

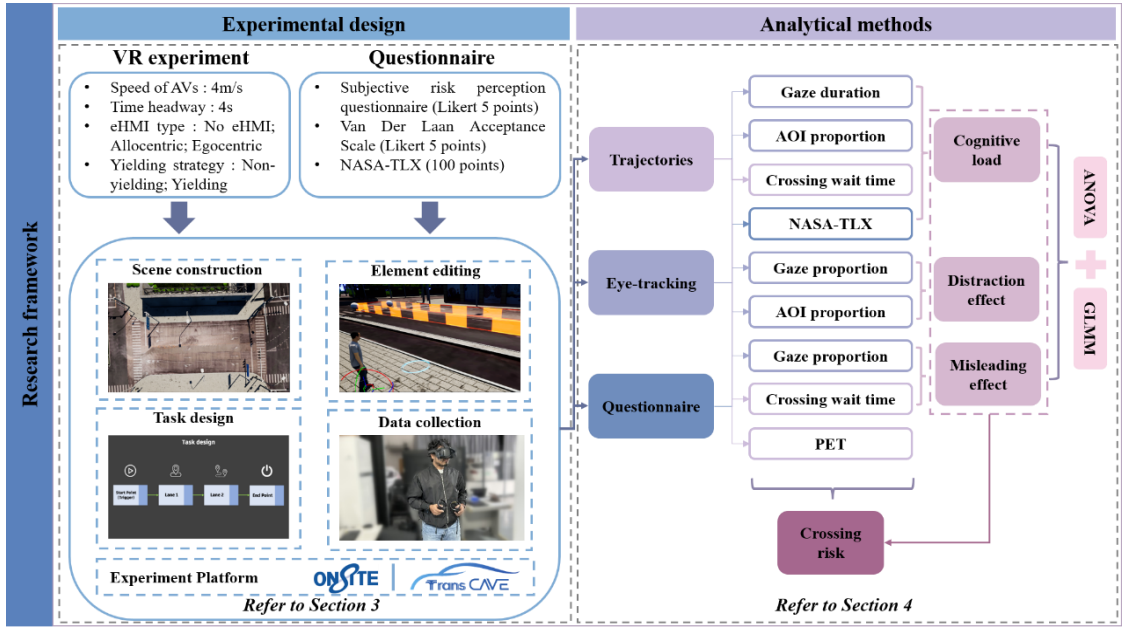

**Fig. 2.** Research framework.

**3.1 Data collection**

*3.1.1 Participants*

A total of 45 participants (31 males and 14 females), aged 20–28 years (mean = 23.29, standard deviation (SD) = 1.74), were recruited for the pedestrian-AV interaction VR experiment, with approval from the university's ethics committee. All participants provided informed consent before the experiment and received compensation upon its completion. Table 2 summarizes the demographic information of the participants.

**Table 2.** Demographic characteristics of participants.

| Variable | Frequency | Proportion |
| --- | --- | --- |
| **Gender** | | |
| Female | 14 | 31.1% |
| Male | 31 | 68.9% |
| **Driver's license holder** | | |
| Yes | 41 | 91.1% |
| No | 4 | 8.9% |
| **Actual driving experience** | | |
| <1 year | 20 | 44.4% |
| 1–3 years | 16 | 35.6% |
| 3–5 years | 7 | 15.6% |
| >5 years | 2 | 4.4% |
| **Educational level** | | |
| Undergraduate | 4 | 8.9% |
| Postgraduate | 41 | 91.1% |
| **Researcher or practitioner in the transportation field** | | |
| Yes | 35 | 77.8% |
| No | 10 | 22.2% |

*3.1.2 Apparatus*

To safely capture AV–pedestrian interaction behaviors during street crossings, a virtual bidirectional four-lane urban pedestrian crossing scenario was created using Unreal, a powerful game development engine (Fig. 3). The participants were required to wear a VIVE Focus 3 headset equipped with eye-tracking functionality and use a controller to control a virtual character, as shown in Fig. 3. The VIVE Focus 3 headset features a resolution of 4896 × 2448 pixels (2448 × 2448 pixels per eye) and a refresh rate of 90 Hz. The experiment was conducted in a spacious laboratory, which allowed the participants to move freely. Once the scenario was initiated, the vehicle traveled from left to right at a constant speed of 4 m/s.

To accurately track the gaze behavior of the participants throughout the experiment and collect comprehensive eye-tracking data for subsequent in-depth analysis, an eye tracker was installed inside the headset. Using heat maps and gaze-tracking functions, this system measured the gaze time and direction of the participants, enabling a comprehensive analysis of vehicle–pedestrian interaction behaviors. The eye tracker provides gaze data for both eyes at a 120 Hz frequency, with an accuracy of 0.5°– 1.1°. It uses a 9-point calibration method, is compatible with both the Unity and Unreal development engines, and can output data such as timestamps, gaze origin, and gaze direction.

The experiment was built and conducted on the TransCAVE platform, owned by the OnSite Committee at Tongji University (https://www.onsite.com.cn). This experimental platform provides a highly customized and immersive dynamic interactive virtual reality environment, features an extensive library of traffic scenarios, and supports the flexible construction of diverse interactive test environments involving pedestrians and traffic systems. This setup was utilized in this study to investigate pedestrian–AV interaction behaviors.

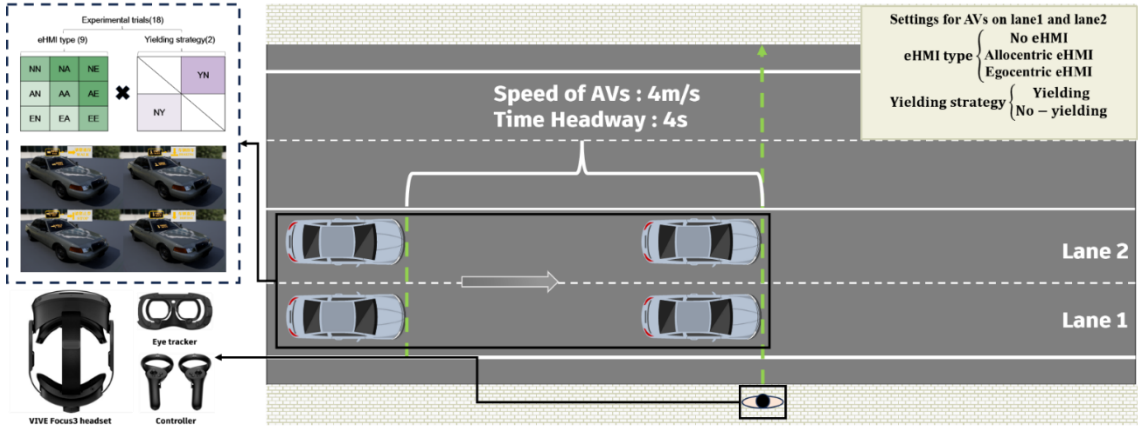

**Fig. 3.** Overview of virtual environment and apparatus (eHMI type: no eHMI (N), allocentric eHMI (A), and egocentric eHMI (E); yielding strategy: yielding (Y) and non-yielding (N)).

*3.1.3 Experimental design*

Regarding the AVs, two key independent variables were considered: (1) the eHMI type based on perspective (i.e., no eHMI, egocentric eHMI, and allocentric eHMI) and (2) the interaction strategy (i.e., yielding or non-yielding behavior). Fig. 3 illustrates the experimental scene and kinematic parameters. Two streams of AVs appeared simultaneously in the two lanes ahead of the pedestrian, with each vehicle traveling at a constant speed of 14.4 km/h and maintaining a headway of 4 s. AVs were continuously generated to ensure that the participants would have to interact with vehicles during the crossing task.

At the combinatorial level, the three types of eHMIs were systematically paired across the two lanes, which resulted in nine distinct eHMI combinations. To accurately capture the complexity of inconsistent vehicle behaviors in real-world multi-agent interactions while also limiting the total number of trials and preserving the representativeness of the key-factor combinations, a fractional factorial design was utilized (Tait et al., 2013). Specifically, in terms of the vehicle interaction strategy, scenarios in which the AVs on the two lanes exhibited different yielding behaviors were selectively included to enhance variability and ecological validity. Consequently, 18 experimental conditions were generated, as summarized in Fig. 3, which also illustrates the presentation of the two types of eHMIs within the experimental scenarios. As this study focused on analyzing the effects of different eHMI perspectives on pedestrians, the eHMIs primarily displayed text-based messages. To eliminate any potential interference caused by color differences, a controlled-variable approach was adopted, whereby all eHMIs displayed their messages in a uniform color.

To examine the crossing behaviors of the participants across various eHMI combinations, a range of independent variables were considered, including demographic factors, experimental design variables, and subjective measures, which are listed in Table 3.

**Table 3.** Descriptions of independent variables.

| Demographic variable | Interpretation | Mean (SD) |
|---|---|---|
| Gender | Male = 0 (68.9%), Female = 1 (31.1%) | — |
| Age | Age of pedestrians, ranging from 20–28 years | 23.289 (1.730) |
| Education | Undergraduate = 0 (8.9%), Postgraduate = 1 (91.1%) | — |
| Practitioner | No = 0 (77.8%), Yes = 1 (22.2%) | — |
| Driver's license holder | No = 0 (8.9%), Yes = 1 (91.1%) | — |
| Actual driving experience | <1 year = 1 (44.4%), 1–3 years = 2 (35.6%), 3–5 years = 3 (15.6%), >5 years = 4 (4.4%) | — |
| **Experiment design variable** | **Interpretation** | **Mean (SD)** |
| Yielding strategy | Non-yielding = 0 (50.0%), Yielding = 1 (50.0%) | — |
| Lane 1 eHMI type | No eHMI = 0 (33.3%), | — |

| | | | | | |
|---|---|---|---|---|---|
| Lane 2 eHMI type | Allocentric = 1 (33.3%), Egocentric = 2 (33.3%) No eHMI = 0 (33.3%), Allocentric = 1 (33.3%), Egocentric = 2 (33.3%) | — | | | |
| Misleading Scenario 1 | Misleading = 1 (33.3%) No misleading = 0 (66.7%) | — | | | |
| Misleading Scenario 2 | Misleading = 1 (33.3%) No misleading = 0 (66.7%) | — | | | |

| Subjective variable | Interpretation | Mean (SD) |
|---|---|---|
| Understanding of AV | 5-point Likert scale | 4.022 (0.174) |
| Trust in AV | 5-point Likert scale | 3.715 (0.305) |
| Risk perception related to AV | 5-point Likert scale | 2.271 (0.338) |
| Understanding of eHMI | 5-point Likert scale | 3.000 (0.097) |
| Trust in eHMI | 5-point Likert scale | 3.844 (0.135) |
| Crossing focus | 5-point Likert scale | 3.517 (0.451) |
| Crossing adherence | 5-point Likert scale | 4.167 (0.083) |
| Attitudes toward others | 5-point Likert scale | 4.341 (0.270) |

| Dependent variable | Interpretation | Mean | SD | Min | Max |
|---|---|---|---|---|---|
| Gaze time for Lane 1 AV | Time of gaze on Lane 1 AV before crossing | 7.516 | 4.511 | 0.225 | 40.450 |
| Gaze proportion for Lane 1 eHMI | Proportion of gaze on Lane 1 eHMI before crossing | 0.163 | 0.131 | 0 | 0.556 |
| Gaze time for Lane 2 AV | Time of gaze on Lane 2 AV before crossing | 0.262 | 0.155 | 0 | 0.791 |
| Gaze proportion for Lane 2 eHMI | Proportion of gaze on Lane 2 eHMI before crossing | 0.126 | 0.172 | 0 | 1.000 |
| Pre-crossing wait time | Wait time before crossing | 11.450 | 5.254 | 2.622 | 41.814 |
| Misleading Behavior 1 | Misleading = 1 (22.5%) No misleading = 0 (77.4%) | — | — | — | — |
| Misleading Behavior 2 | Misleading = 1 (27.4%) | — | — | — | — |
| Conflict in Lane 1 | No misleading = 0 (72.5%) Serious conflict = 1 (49.2%) Non-serious conflict = 0 (50.7%) | — | — | — | — |
| Conflict in Lane 2 | Serious conflict = 1 (48.3%) Non-serious conflict = 0 (51.6%) | — | — | — | — |

*3.1.4 Questionnaire design*

**(1) Measurement of presence**

The ITC-Sense of Presence Inventory (ITC-SOPI) questionnaire was designed by Lessiter et al. (2001) to measure a participant's subjective sense of presence in a VR environment across four dimensions: spatial presence, engagement, ecological validity, and negative effects.

- **Spatial presence** assesses the participant's sense of being in the presented environment, based on statements such as "I feel like I am in the displayed scene" and "I feel that the characters and/or objects are almost touching me."
- **Engagement** evaluates psychological involvement and enjoyment during the experience, based on statements such as "I feel like I am involved," "I enjoyed myself," and "My experience was intense."
- **Ecological validity** refers to the credibility and naturalness of the environment, based on statements such as "The environment looks very natural" and "The characters and objects look very realistic."
- **Negative effects** measure any adverse physiological reactions that participants may experience, based on statements such as "I felt dizzy," "I felt nauseous," and "My eyes felt tired."

The participants were required to complete the ITC-SOPI questionnaire at the end of the VR experiment.

**(2) Measurement of simulator sickness**

The Simulator Sickness Questionnaire (SSQ) is widely used for assessing simulator sickness (Bouchard et al., 2007). It quantifies symptoms induced by virtual environments and evaluates their severity across three main symptom clusters:

- **Oculomotor symptoms**: eye fatigue, difficulty in focusing, blurred vision, and headaches;
- **Disorientation**: dizziness and vertigo;
- **Nausea-related symptoms**: nausea, stomach discomfort, increased salivation, and burping.

Each symptom cluster is scored separately to ensure a comprehensive assessment. The participants were required to complete this questionnaire after every four VR experiment trials to ensure an accurate evaluation of simulator sickness.

**(3) Measurement of cognitive load**

The NASA Task Load Index (NASA-TLX), developed by the National Aeronautics and Space Administration (NASA), is one of the most widely used tools for assessing subjective psychological workload. It exhibits minimal between-subject variability and has gained wide user acceptance (Febiyani et al., 2021). Table A1 presents its design and scoring criteria.

The NASA-TLX questionnaire evaluates workload across six dimensions, namely Mental Demand (MD), Physical Demand (PD), Temporal Demand (TD), Own Performance (OP), Effort (EF), and Frustration Level (FR), each of which is scored on a 100-point scale. After completing the subjective ratings, participants compare all six dimensions in pairwise combinations and select the more important one in each pair, generating a count of selections per dimension, as summarized in Table A3. The weighted score is then calculated using Eq. (1):

$$F = \sum_{i=1}^{6} M_i \times \frac{P_i}{15} \tag{1}$$

where $F$ represents the total cognitive workload score, $M_i$ is the participant's score on the $i$-th dimension, and $P_i$ is the number of times the $i$-th dimension was selected in the weight comparison table (Table A3).

**(4) Measurement of attitudes**

The Van Der Laan scale (Van Der Laan et al., 1997) is widely used in scenarios involving automation systems, driver assistance systems, and AVs to help researchers assess the emotional responses, acceptance, and expectations of users toward new technologies. It consists of four dimensions, each measured using a Likert scale (e.g., 1–5 points) to evaluate the attitudes and perceptions of users:

- **Perceived usefulness**: It assesses whether users believe a technology or system enhances efficiency, comfort, or other functionalities.
- **Perceived ease of use**: It measures how easy users find the technology to use.
- **Trust**: It evaluates the level of trust users have in the technology, a key factor in its acceptance.
- **Satisfaction**: It gauges overall user satisfaction as an indicator of acceptance.

The participants were required to complete this questionnaire at the end of the VR experiment.

**(5) Measurement of risk perception**

The subjective risk perception questionnaire used in this study measures a participant's understanding of AVs and eHMIs, as well as their risk perception, on a 5-point Likert scale across nine dimensions: demographic background, understanding of AVs (Li et al., 2024), trust in AVs, perception of risk from AVs (Deb et al., 2017), understanding of eHMIs, trust in eHMIs (Carmona et al., 2021), the participant's own attention levels while crossing the street, the participant's adherence to crossing norms, and the participant's attitudes toward other road users (Esmaili et al., 2021).

The participants were required to complete this questionnaire at the end of the VR experiment.

To measure risk perception, latent variable scores were estimated through confirmatory factor analysis (CFA), based on factor scores derived from the observed questionnaire items. These latent scores were then incorporated into analytical models. The main steps for calculating the latent variable scores are as follows:

- Standardized factor loadings ($\lambda$) for each observed variable were obtained using the partial least-squares structural equation modeling algorithm (Sarstedt et al., 2022).
- Each observed variable was then converted into a Z-score, which is a standardized score representing the deviation of a raw score from the mean, expressed as an SD. It is calculated using Eq. (2):

$$Z = \frac{(X - \mu)}{\sigma} \tag{2}$$

where $X$ is the original score, $\mu$ is the mean, and $\sigma$ is the SD. Finally, the latent variable scores were calculated using Eq. (3):

$$\xi = \sum (\lambda \times Z) \tag{3}$$

where $\xi$ is the latent variable score, $\lambda$ is the factor loading coefficient, and $Z$ is the Z-score of the corresponding observed variable.

### 3.1.5 Experimental procedure

Upon arriving at the laboratory, the participants were first asked to sign an informed consent form detailing the potential risks and discomfort. The form explained that participants might experience the type of stress caused by approaching vehicles in real life, although this stress would not exceed normal levels. If they experienced any discomfort, they could inform the experimenter and stop the process immediately. Additionally, mild fatigue or discomfort could occur, and the participants were free to rest or withdraw from the VR experiment at any time.

The experiment involved video recordings and questionnaires, with all collected information kept strictly confidential. Fig. 4 presents the overall experimental procedure, which included the following steps:

**Step 1. Introduction to the experiment**

The experimenter described the overall procedure and provided the key safety instructions. The participants were informed that they would decide when to cross the street based on their own safety judgment and should pay attention to any information displayed on the roof, windshield, or sides of the vehicles, which might convey vehicle status or provide crossing guidance. They were encouraged to interpret this information in context and cross the street safely. The experimenter then provided detailed instructions regarding the usage of the VR equipment and handheld controller.

**Step 2. Eye-tracking calibration**

After familiarizing themselves with the experimental procedure and device operation, the participants were assisted in wearing the VR headset. To ensure accurate data collection, they were required to complete the calibration process for the built-in eye-tracking system.

**Step 3. Pre-experiment**

Once all questions had been addressed, the participants completed a brief pre-experiment to become accustomed to the procedure. In this scenario, vehicles appeared only in a single lane, and all of them were equipped with eHMIs, which allowed the participants to intuitively recognize the different eHMI types. After the pre-experiment, the participants completed the SSQ to ensure that their adaptation to the VR environment could be assessed. The pre-experiment comprised two trials, with additional trials conducted if deemed necessary based on participant performance.

**Step 4. Formal experiment**

The formal experiment comprised 18 trials. After each trial, the subjective cognitive load on the participants due to the crossing event was assessed based on the NASA-TLX scale. To monitor the simulator sickness experience by the participants, the SSQ was administered every four trials.

At the end of the experiment, the influence of different eHMI combinations on the crossing decisions of the participants was assessed based on the Van Der Laan scale, their subjective sense of presence in the VR environment was evaluated based on the ITC-SOPI questionnaire, and their understanding of the AVs and eHMIs, as well as their perceived risk levels, was measured based on the subjective risk perception questionnaire.

**Step 5. End of experiment**

After completing the questionnaires, the participants were provided with monetary compensation and informed that the experiment had concluded.

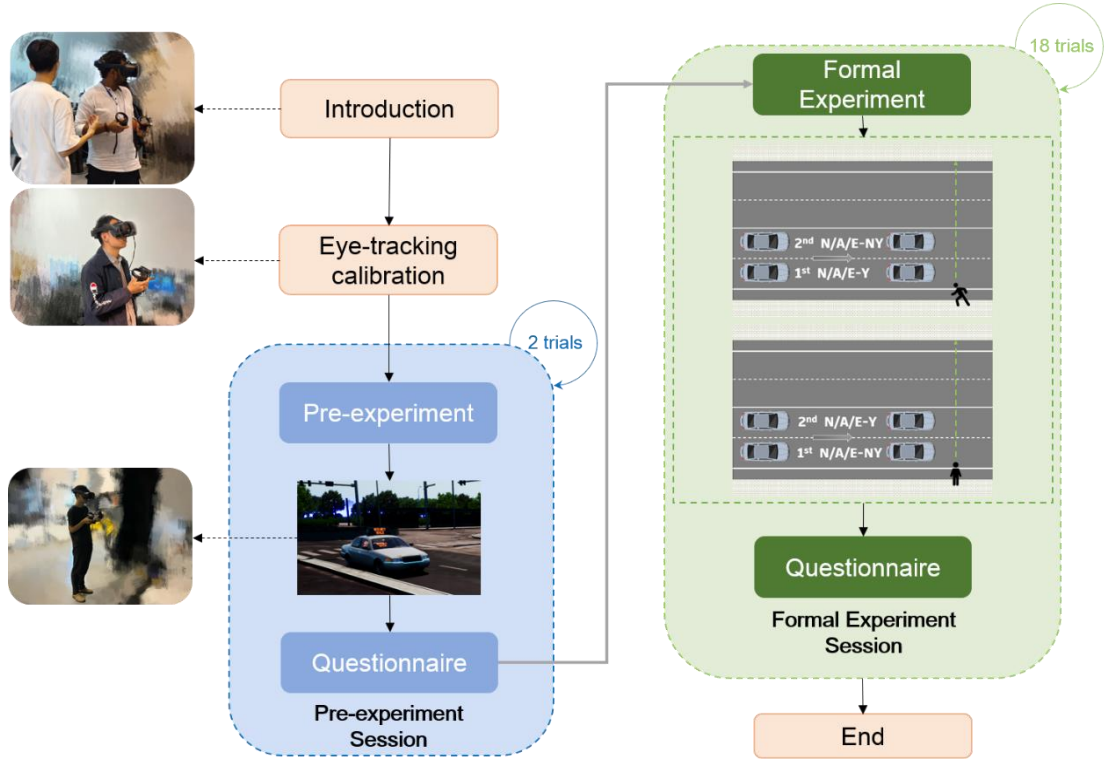

**Fig. 4.** Overall experimental procedure.

## 3.2 Statistical analysis

*3.2.1 Cognitive load*

Cognitive load theory provides a key perspective for understanding how easily pedestrians process eHMI information from AVs. Proposed by Sweller (2011), the theory suggests that human working memory provides a limited capacity for processing information and that the complexity of learning and decision-making tasks influences cognitive load. As task complexity increases, the burden on working memory also rises, which potentially impacts learning and decision-making processes.

In an autonomous driving environment, when pedestrians are faced with eHMI information from multiple AVs, they must simultaneously process visual, cognitive, and contextual information; this may increase cognitive load, thereby affecting the interaction behaviors of pedestrians, such as gaze time and waiting time. Fig. 5(a) presents a heatmap visualizing the gaze patterns of the pedestrians during the experiment.

The cognitive load analysis involved using three primary indicators to measure the impact of different eHMI combinations on the behavior of the pedestrians during crossing:

- **Gaze time for Lane 1 AV:** the time of the gaze toward the AV in Lane 1 before crossing;
- **Gaze proportion for Lane 1 eHMI:** the proportion of the gaze directed toward the eHMI on a Lane 1 AV before crossing, as defined in Eq. (4) and illustrated in Fig. 5(b):

$$P_{eHMI1} = \frac{T_{eHMI1}}{T_{eHMI1} + T_{AV1}} \quad (4)$$

where $T_{eHMI1}$ denotes the gaze time for the eHMI on a Lane 1 AV, and $T_{AV1}$ denotes the gaze time for the AV in Lane 1;
- **Pre-crossing wait time:** the time that the pedestrian waits before crossing;
- **NASA-TLX:** Evaluated subjective cognitive load of pedestrians.

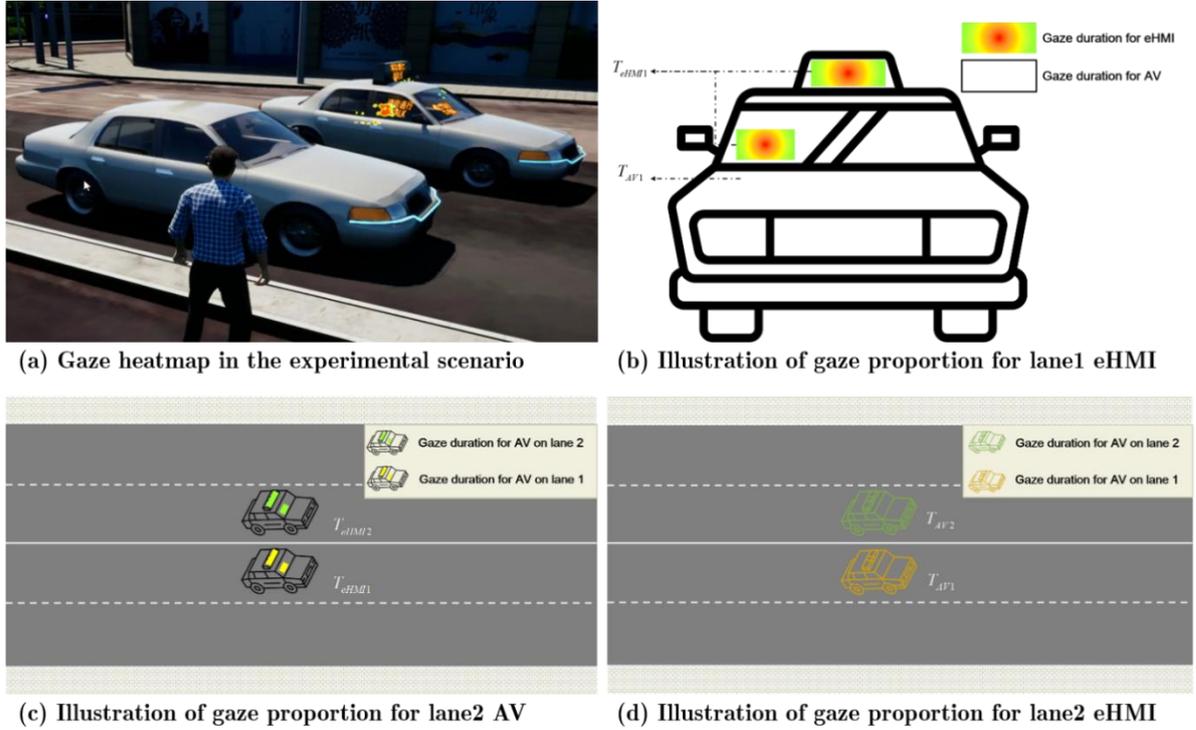

**Fig. 5.** Calculation of cognitive load and distraction effect: (a) Gaze heatmap during experiment; (b) Gaze proportion for Lane 1 eHMI; (c) Gaze proportion for Lane 2 AV; (d) Gaze proportion for Lane 2 eHMI.

*3.2.2 Distraction effect*

When pedestrians cross the street, excessive visual objects in the scene can lead to improper attention allocation during tasks that rely primarily on visual input. As the main decision-making process during pedestrian street crossing typically occurs in the pre-crossing phase, this section uses two indicators to measure the effect of distractions on crossing decisions made by pedestrians:

- **Gaze proportion for Lane 2 AV:** the proportion of the gaze toward the AV in Lane 2 before crossing, which is calculated using Eq. (5) and illustrated in Fig. 5(c):

$$P_{AV2} = \frac{T_{AV2}}{T_{AV1} + T_{AV2}} \quad (5)$$

where $T_{AV1}$ and $T_{AV2}$ denote the gaze time for the AVs in Lanes 1 and 2, respectively.

- **Gaze proportion for Lane 2 eHMI:** the proportion of the gaze toward the eHMI on a Lane 2 AV before crossing, which is determined using Eq. (6) and illustrated in Fig. 5(d):

$$P_{eHMI2} = \frac{T_{eHMI2}}{T_{eHMI1} + T_{eHMI2}} \quad (6)$$

where $T_{eHMI1}$ and $T_{eHMI2}$ denote the gaze time for the eHMIs on the AVs in Lanes 1 and 2, respectively.

*3.2.3 Misleading effect*

In complex, multi-lane traffic environments, multi-source eHMI information can not only cause distraction but also transmit ambiguous or conflicting signals that may mislead the crossing decisions of pedestrians. This study focuses specifically on misleading effects associated with potential risks, which are categorized into two main potential misleading scenarios:

- **Misleading Scenario 1:** Misleading information from the eHMI on a Lane 1 AV instructs the pedestrian to cross the street, causing them to place excessive trust in this eHMI. Thus, the pedestrian may pay insufficient attention to AVs in Lane 2 and may not wait long enough when interacting with these AVs.

- **Misleading Scenario 2:** Misleading information from the eHMI on a Lane 2 AV instructs the pedestrian to cross the street, leading the pedestrian to over-rely on this eHMI. This causes the pedestrian to not focus adequately on AVs in Lane 1 before crossing and to not wait long enough when interacting with these AVs.

Typical examples of these potential misleading scenarios are presented in Fig. 6(a) and 6(b), while Fig. 6(c) shows all possible combinations of misleading scenarios.

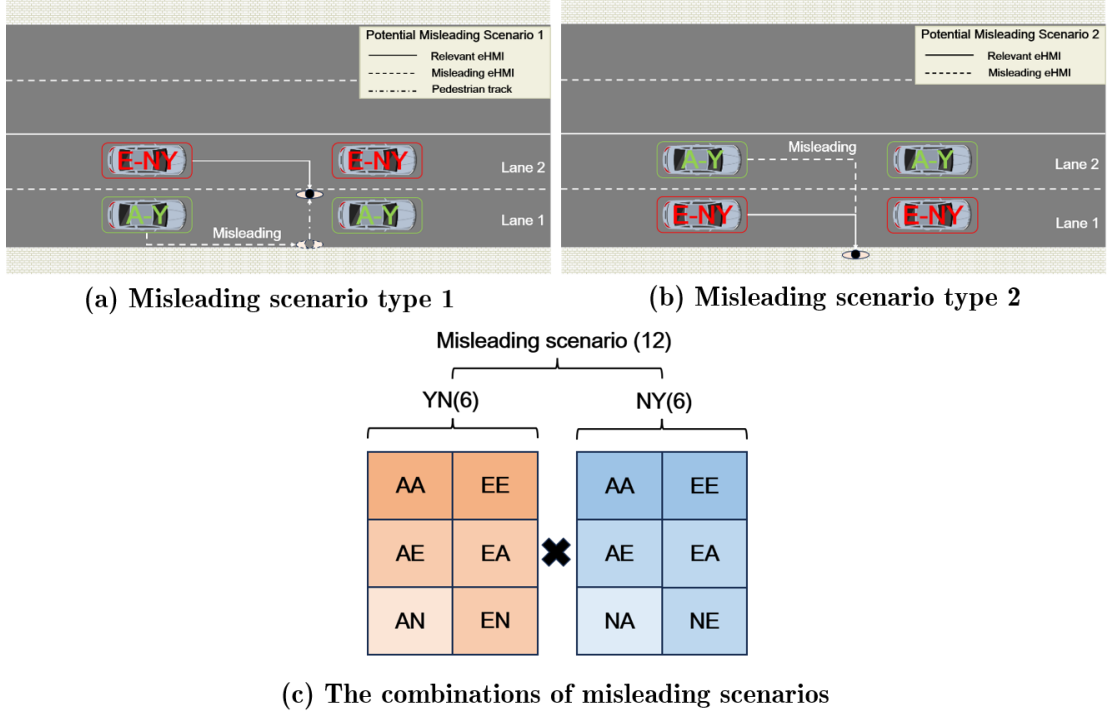

(a) Misleading scenario type 1

(b) Misleading scenario type 2

(c) The combinations of misleading scenarios

**Fig. 6**. Multi-lane pedestrian crossing scenarios with possible misleading effects: (a) **Misleading Scenario 1**; (b) **Misleading Scenario 2**; (c) Combinations of misleading scenarios (eHMI type: no eHMI (N), allocentric eHMI (A), and egocentric eHMI (E); yielding strategy: yielding (Y) and non-yielding (N)).

Regarding the two types of misleading scenarios described above, the criteria for determining whether a pedestrian was actually misled during the crossing process are as follows:

- **Misleading Behavior 1:** When crossing Lane 2, if the duration of the pedestrian's gaze toward the AVs in Lane 2 and their waiting time before crossing Lane 2 fall below certain minimum thresholds, the pedestrian is considered to have been misled by Misleading Scenario 1.

- **Misleading Behavior 2:** Before crossing, if the pedestrian's gaze time toward the vehicles in Lane 1 and their pre-crossing waiting time fall below certain minimum thresholds, the pedestrian is considered to have been misled by Misleading Scenario 2.

Outlier detection is a common approach to setting thresholds for key indicators. However, techniques based on conventional outlier detection parameters, such as SD, median absolute deviation, and interquartile range (IQR), rely heavily on the statistical properties of the data. Thus, these methods are often sensitive to the presence of extreme outliers, which can distort the computed thresholds and hinder the accurate identification of true outliers. To enhance the robustness of threshold determination, the two-stage thresholding (2T) method proposed by Yang et al. (2019) is adopted. This approach, based on the traditional IQR technique, improves the accuracy of outlier detection by reducing the influence of extreme values on threshold computation.

In the first stage, the IQR was calculated to identify extreme values that exceeded the preliminary threshold, as defined in Eq. (7):

$$\text{Upperbound} = Q_3 + 1.5 \times IQR, \ \text{Lowerbound} = Q_1 - 1.5 \times IQR \tag{7}$$

All observations falling outside this range were temporarily excluded.

In the second stage, the IQR was re-calculated from the cleaned dataset (with extreme values removed), and the final thresholds were determined using the updated Q1 and Q3 values. This two-step process ensures that the final thresholds are less affected by extreme outliers and reflect the distribution of the main data body more accurately.

To establish reasonable thresholds for identifying potentially misleading behaviors, scenarios without misleading conditions were first selected as the baseline group. Based on these normal crossing conditions, the typical distribution ranges of the key behavioral indicators (e.g., gaze time and waiting time) were determined using the 2T method when the pedestrians were not misled. The derived thresholds were then employed as the criteria to assess whether pedestrians were actually misled in experimental scenarios involving potential misleading factors.

Misleading Behavior 1 involved the pedestrians interacting with non-yielding vehicles in Lane 2 and being potentially misled by the eHMI signals displayed by yielding vehicles in Lane 1. Therefore, three scenarios in which no eHMI was present on the Lane 1 AVs were selected as the baseline group (including combinations where Lane 2 AVs were equipped with no eHMI, egocentric eHMIs, or allocentric eHMIs) to construct the reference threshold, as illustrated in Fig. 7(a). The occurrence of misleading was determined by examining whether the gaze time and waiting time of the pedestrians during their interactions with the Lane 2 vehicles fell below the threshold, as formulated in Eq. (8):

$$\text{Mislead}_{\text{type1}} = \begin{cases} 1, & \text{if } T_{gaze\_AV2} < \theta_{gaze}^{(1)} \text{ and } T_{wait\_AV2} < \theta_{wait}^{(1)} \\ 0, & \text{otherwise} \end{cases} \quad (8)$$

where $T_{gaze\_AV2}$ refers to the duration of the gaze directed toward the vehicle in Lane 2, $T_{wait\_AV2}$ denotes the waiting time before interacting with the vehicle in Lane 2, and $\theta_{gaze}^{(1)}$ and $\theta_{wait}^{(1)}$ are the corresponding thresholds derived from the baseline scenarios in which no eHMI was present on the Lane 1 AVs.

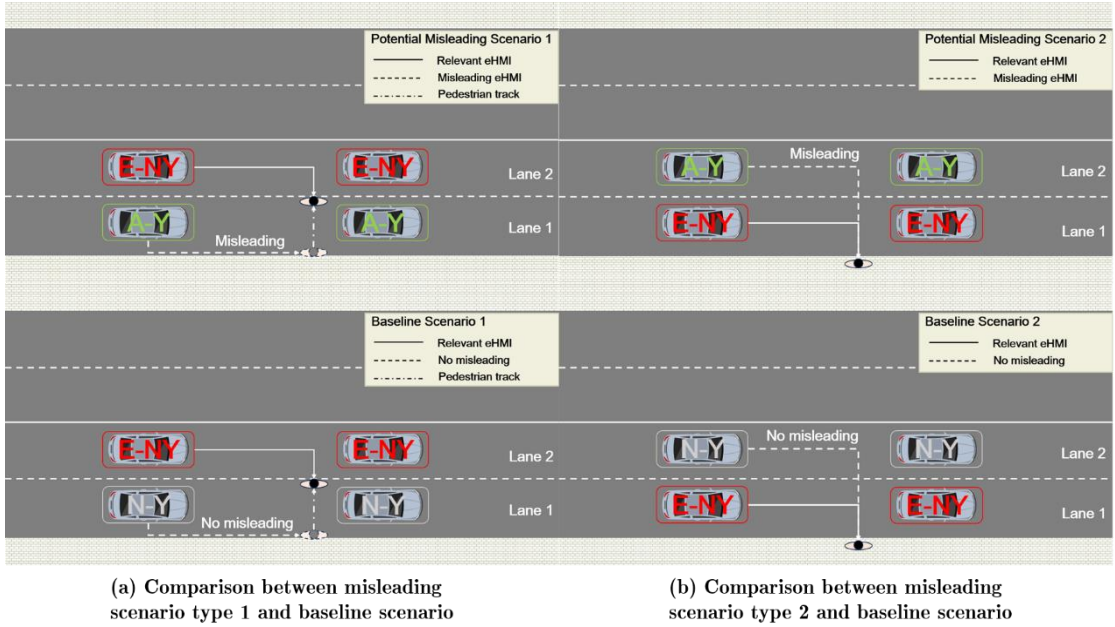

**Fig. 7.** Comparison between potential misleading scenarios and baseline scenarios for threshold extraction: (a) Misleading Scenario 1; (b) Misleading Scenario 2.

Misleading Behavior 2 involved the pedestrians interacting with non-yielding vehicles in Lane 1 and being potentially misled by the eHMI information provided by yielding vehicles in Lane 2. Accordingly, three scenarios where no eHMI was present on the Lane 2 AVs were selected as the baseline group (including combinations where the Lane 1 AVs were equipped with no eHMI, egocentric eHMIs, or allocentric eHMIs), as illustrated in Fig. 7(b). The occurrence of misleading was determined by assessing whether the gaze time and waiting time for the Lane 1 vehicles fell below the respective established thresholds, as formulated in Eq. (9):

$$\text{Mislead}_{type2} = \begin{cases} 1, \text{ if } T_{gaze\_AV1} < \theta_{gaze}^{(2)} \text{ and } T_{wait\_AV1} < \theta_{wait}^{(2)} \\ 0, \text{ otherwise} \end{cases} \quad (9)$$

where $T_{gaze\_AV1}$ refers to the duration of the gaze directed toward the vehicle in Lane 1, $T_{wait\_AV1}$ denotes the waiting time before crossing, and $\theta_{gaze}^{(2)}$ and $\theta_{wait}^{(2)}$ are the corresponding thresholds derived from the baseline scenarios in which no eHMI was present on the Lane 2 AVs.

*3.2.4 Crossing risk*

In the context of this study, "crossing risk" specifically refers to the risk of a potential collision between a pedestrian and a vehicle. This risk is measured based on post encroachment time (PET), which is defined as the time interval between a pedestrian leaving the potential collision area and the vehicle reaching the potential collision point (Cooper, 1984; Howlader et al., 2024). PET was calculated from the experimental data using Eq. (10), where $T_1$ represents the time point when the pedestrian leaves the potential conflict area, and $T_2$ represents the time point when the vehicle arrives at the potential conflict area.

$$PET = T_2 - T_1 \quad (10)$$

To categorize the calculated risk levels, the PET values were classified into binary categories based on established thresholds (Islam et al., 2023). Specifically, a PET of less than 1.5 s was defined as a serious conflict (designated as 1), while a PET greater than or equal to 1.5 s was defined as a non-serious or safe interaction (designated as 0); this is because a PET threshold of 1.5 s corresponds to the strongest correlation between crashes and conflicts (Zheng et al., 2016; Islam et al., 2023). This binary risk indicator was subsequently used in the GLMM analysis to examine the factors contributing to high-risk AV–pedestrian interactions.

*3.2.4 Generalized linear mixed-effects model*

A GLMM was chosen to analyze the variations in different dependent variables across participants as it not only allows for the simultaneous estimation of fixed effects (e.g., the impact of different eHMI combinations on pedestrian gaze behavior) and random effects (e.g., individual differences among participants) but also supports the modeling of dependent variables with different distributions (e.g., binomial and Poisson distributions) by applying appropriate link functions to establish relationships between independent and dependent variables. Thus, it represents a highly flexible approach capable of handling various types of dependent variables and modeling data with a hierarchical structure. The GLMM is defined in Eq. (11):

$$g\left(\mathbb{E}\left[Y_{ij}\right]\right) = X_{ij}\beta + Z_{ij}u_j \tag{11}$$

where $Y_{ij}$ represents the dependent variable in the $i$-th observation of the $j$-th group of random effects (e.g., different individuals and different experimental groups), and $g(\bullet)$ is the link function, which maps the expectation $\mathbb{E}\left[Y_{ij}\right]$ to the linear prediction space. The specific form of the link function depends on the distribution of the response variable. For example, the logit function is used for binomial distributions, while the log function is used for Poisson distributions. Additionally, $\mathbb{E}\left[Y_{ij}\right]$ is the expectation of $Y_{ij}$; $X_{ij}$ is the design matrix for fixed effects, which includes the values of fixed-effect variables, such as covariates; $\beta$ is the fixed-effect coefficient (regression coefficient), which represents the effect of the independent variable on the response variable; $Z_{ij}$ is the design matrix for random effects, which describes the random-effect variables; and $u_j$ is the random-effect term, which captures random variations associated with individuals or experimental groups and is typically assumed to follow a normal distribution $N(0, \sigma^2)$.

In this study, the behavioral and physiological indicators were used as the dependent variables, while the experimental design variables, demographic variables, and subjective variables were treated as the independent variables. Each participant was modeled as a random effect to account for individual variability.

If the scenario variable was found to be statistically significant in any GLMM, post hoc analyses were conducted to further examine the differences between conditions. Pairwise comparisons were performed among different experimental combinations, and Bonferroni correction was applied to adjust for multiple comparisons. The contrast coefficients derived from the post hoc tests represent the estimated differences in effects between conditions, which enabled the identification of specific combinations that differed significantly. All statistical analyses were conducted at a 95% confidence level (α = 0.05).

## 4. Results

### 4.1 Immersiveness and simulator sickness

To ensure the reliability of the VR experiment results, the sense of immersion experienced by the participants was measured using the ITC-SOPI scale (Lessiter et al., 2001), with scores ranging from 1 to 5 (Table 4).

Table 4. Results of ITC-SOPI analysis.

| ITC-SOPI score | Mean (SD) score for different aspects | | | |
|---|---|---|---|---|
| | Spatial presence | Engagement | Naturalness | Negative effects |
| | 3.69 (0.54) | 3.56 (0.47) | 3.67 (0.65) | 2.11 (0.74) |

The mean scores for spatial presence, engagement, and naturalness were all above 3.0, while the scores for negative effects remained below 2.50. These results indicate that the participants perceived a high level of presence within the VR simulator environment.

Additionally, the discomfort of the participants after each experiment was assessed through the SSQ. Based on the categorization of symptoms outlined by Bouchard et al. (2007), the classification criteria are shown in Table 5. Table 6 and Fig. 8 summarize the overall results.

Table 5. Categorization of SSQ symptoms.

| Score | Category |
|---|---|
| **0** | No symptoms |
| **<5** | Negligible symptoms |
| **5–10** | Minimal symptoms |
| **10–15** | Significant symptoms |
| **15–20** | Concerning symptoms |
| **>20** | Problematic simulator |

Table 6. SSQ recordings throughout experiment.

| SSQ score (F-statistic, $p$-value) | Mean (SD) scores at different times (cumulative virtual environment time) | | | | |
|---|---|---|---|---|---|
| | Scenes 1–2 (warm-up) | Scenes 3–6 | Scenes 7–10 | Scenes 11–14 | Scenes 15–18 |
| **Nausea** | 1.69 (5.49) | 1.48 (4.53) | 2.54 (8.48) | 3.18 (8.87) | 2.54 (7.44) |
| ($F = 0.42$, $p = 0.79$) | | | | | |
| **Oculomotor** | 3.71 (8.02) | 4.55 (11.13) | 7.92 (12.82) | 6.91 (8.97) | 8.59 (9.08) |
| ($F = 1.97$, $p = 0.10$) | | | | | |
| **Disorientation** | 6.19 (18.19) | 6.19 (16.67) | 6.81 (16.95) | 4.95 (9.46) | 5.88 (10.49) |
| ($F = 0.09$, $p = 0.98$) | | | | | |
| Total Score | 4.15 (10.43) | 4.48 (10.92) | 6.73 (12.39) | 5.98 (7.71) | 6.82 (7.58) |
| ($F = 0.70$, $p = 0.59$) | | | | | |

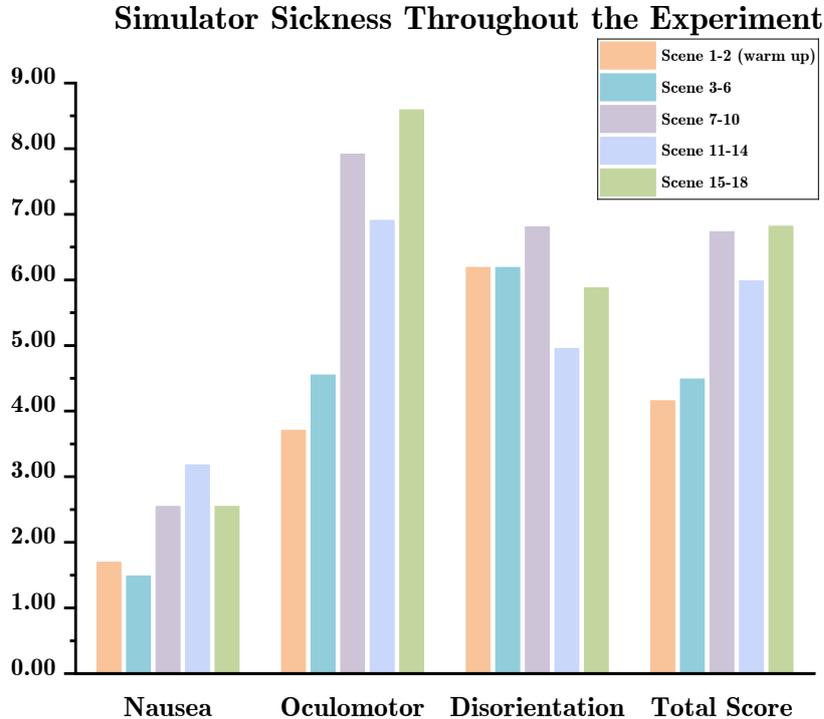

**Fig. 8.** Simulator sickness throughout the experiment.

All participants' SSQ scores in VR remained below 10, indicating minimal symptoms (see Table 6). Simulator sickness slightly increased over time (Fig. 8), but the change was not statistically significant. In the statistical analysis of the SSQ results, the F-statistic reflects between-group differences, and the $p$-value assesses their statistical significance.. A $p$-value of less than 0.05 is typically considered to indicate statistical significance. In the experiment, the $p$-values for all dimensions were well above 0.05, which suggests that the changes in the SSQ scores across different stages were not statistically significant. Overall, the virtual experimental environment demonstrated high usability, and the experimental results are considered reliable.

**4.2 Subjective risk perception**

*4.2.1 Attitudes of participants toward AVs*

The detailed subjective questionnaire is presented in Appendix B. Fig. 9 shows the observations regarding the attitudes of the participants toward AVs. The section on "Understanding of AVs" consists of four items. On average, 78% reported high AV understanding. "Trust in AVs" comprises six items; 81.5% believed AVs improve safety and intended to use them in future travel (Trust in AVs 1 and 2).

However, regarding the interaction behavior and safety of AVs during their operation, only 36% of the participants reported fully trusting AVs ("Trust in AVs" 3 and 5). In terms of willingness to recommend AVs to others, 45.5% of the participants strongly agreed.

The section on "Risk Perception of AVs" includes five items. 82.5% were unconcerned about AV malfunctions or interference with human-driven vehicles ("Risk Perception of AVs" 1–3). However, participants were more cautious about AVs interacting with pedestrians or non-motorized vehicles ("Risk Perception of AVs" 4 and 5). 42% were not concerned about AV–human interaction risks, while 30% were.

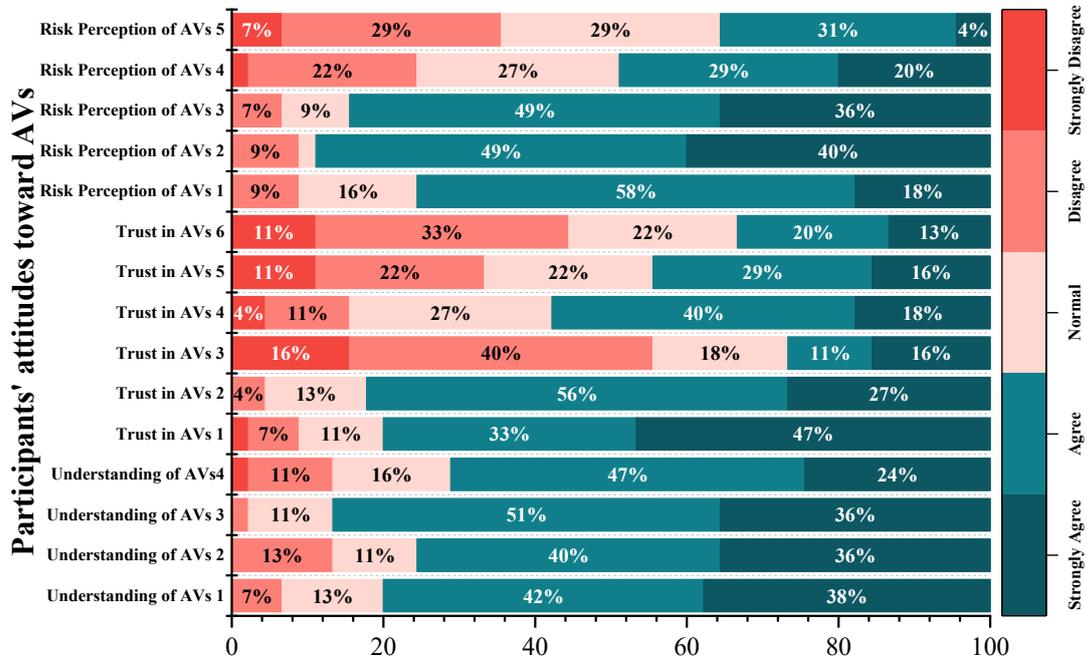

**Fig. 9.** Attitudes of participants toward AVs.

4.2.2 Attitudes of participants toward eHMIs

Fig. 10 presents participant attitudes toward eHMIs. Only 38% reported understanding eHMIs or having prior exposure. However, 72% showed trust in eHMIs' role in safety and accident reduction, indicating future usage willingness ("Trust in eHMI" 1–4).

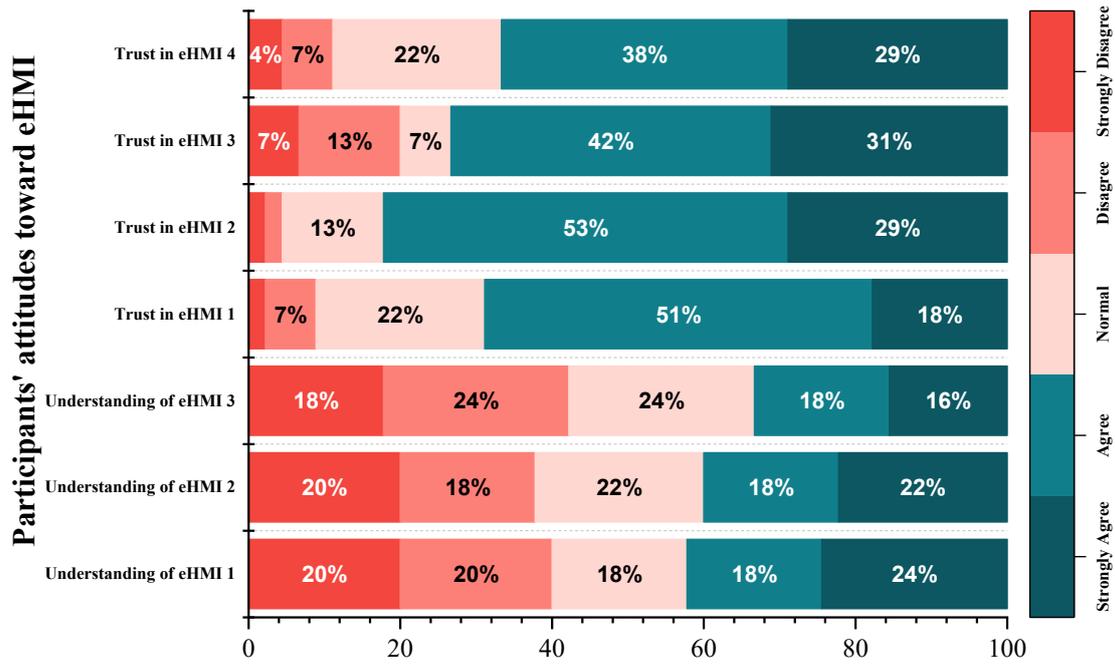

**Fig. 10.** Attitudes of participants toward eHMIs.

*4.2.3 Attitudes of participants toward other road users*

Fig. 11 presents participant attitudes toward other road users. For "Crossing Focus" (4 items), 56% admitted to distractions like chatting or listening to music while crossing, suggesting common inattentiveness. For "Crossing Compliance" (6 items), 78% claimed to follow traffic rules strictly. Regarding interactions, 87% reported a positive attitude (Items 1–4), and 61.5% preferred conservative yielding (Items 5–6).

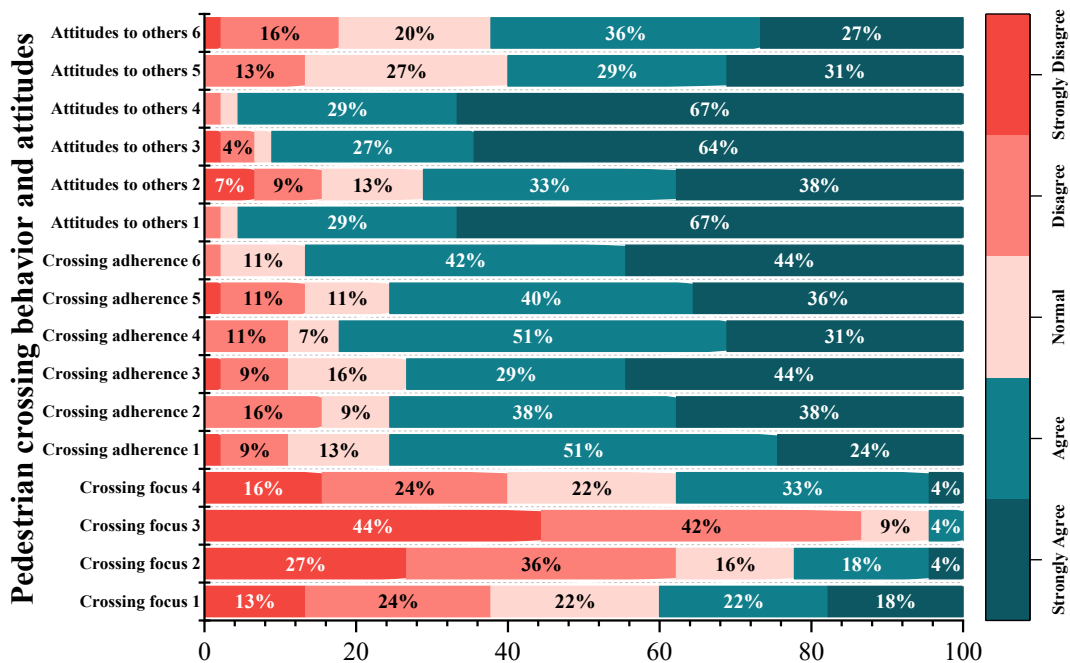

**Fig. 11.** Crossing behaviors and attitudes of pedestrians.

*4.2.4 CFA*

To incorporate the subjective variables into the GLMM as dependent variables, the latent scores for each subjective variable were calculated via CFA. Fig. 12 presents the factor loadings and other statistical indicators for all variables involved.

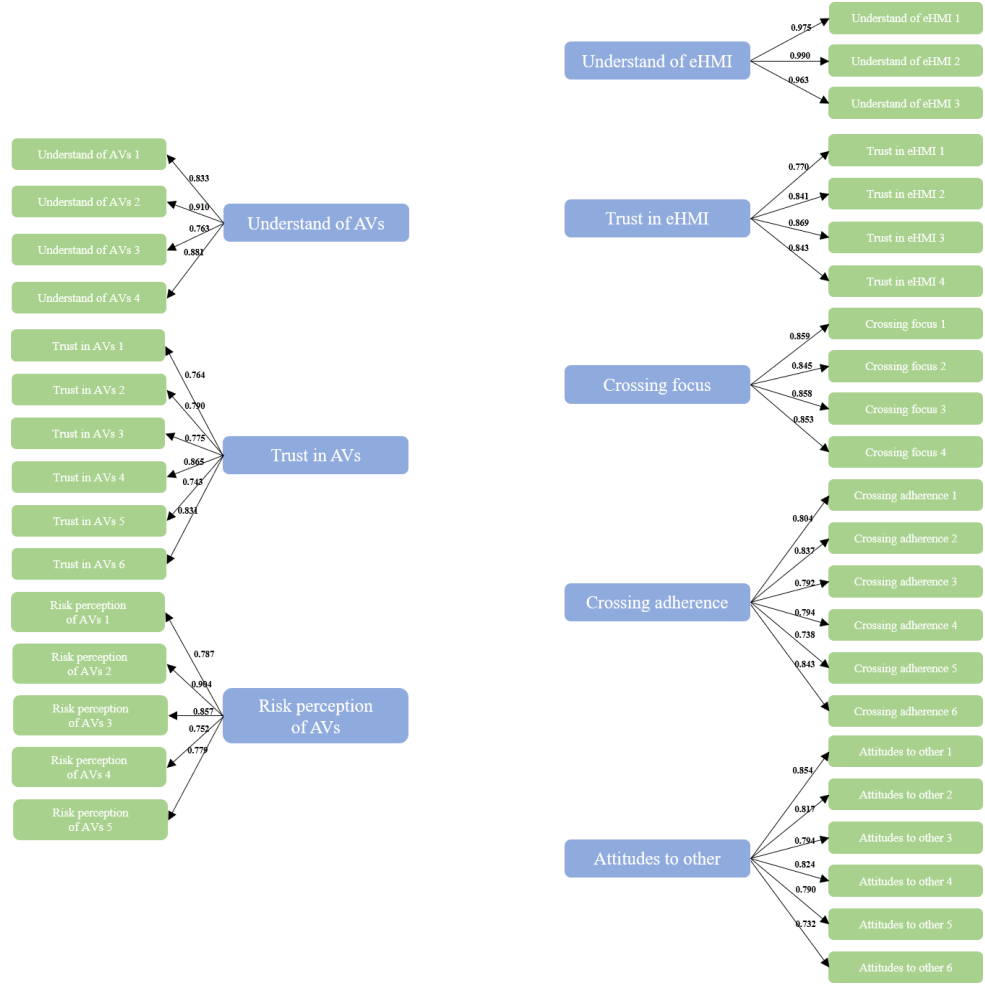

**Fig. 12.** Outer model representation of subjective constructs.

*4.2.5 Van Der Laan acceptance scale*

The Van Der Laan acceptance scale scores for instances of allocentric and egocentric communication were compared to assess the subjective evaluations of the two eHMI types provided by the participants. The results in Fig. 13 indicate that the participants rated egocentric communication higher than allocentric communication in terms of both usefulness and satisfaction.

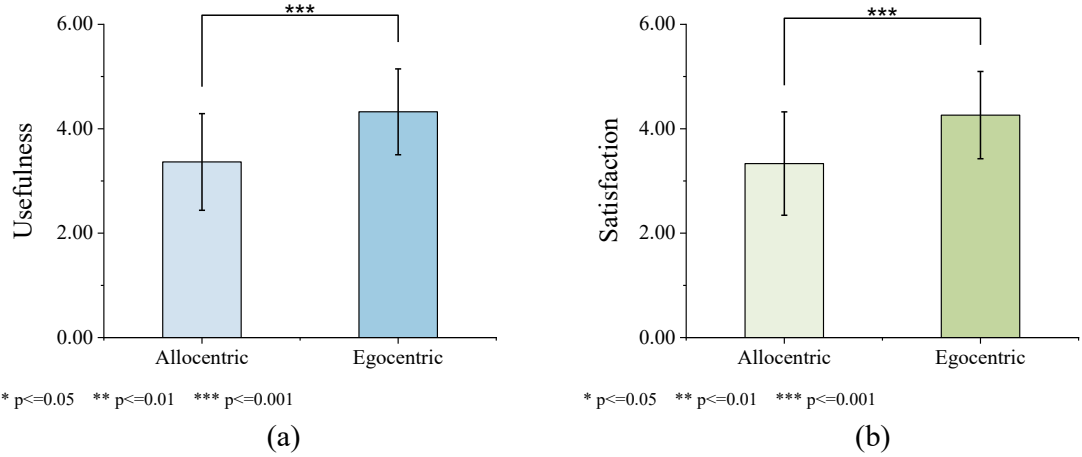

* p<=0.05    ** p<=0.01    *** p<=0.001

(a)                (b)

**Fig. 13.** Van Der Laan acceptance scale scores, with error bars representing the standard error: (a) Usefulness; (b) Satisfaction.

### 4.3 Cognitive load

*4.3.1 Cognitive load across different eHMI types*

In the VR experiment, the eye-tracking area of interest covered AVs and their eHMIs. Heatmaps were generated from aggregated gaze data to assess how different eHMI types affect cognitive load. As shown in Fig. 14(a) and 14(b), when only Lane 1 featured an eHMI-equipped AV, allocentric eHMIs elicited stronger gaze intensity than egocentric ones. This suggests that allocentric communication may impose a higher cognitive load on pedestrians.

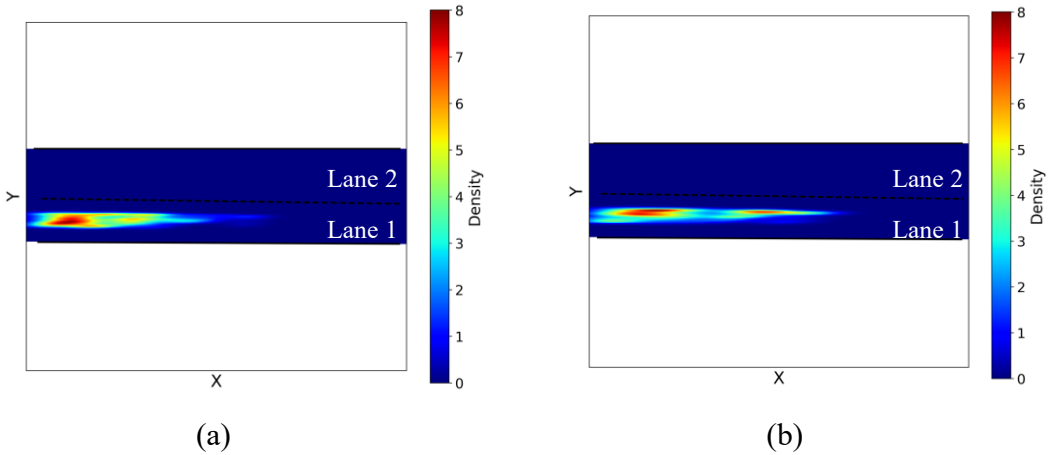

(a)                (b)

**Fig. 14.** Kernel density heatmap of pedestrian gaze on eHMI in different scenarios: (a) Lane 1 with allocentric communication and Lane 2 without eHMI; (b) Lane 1 with egocentric communication and Lane 2 without eHMI.

Based on the experimental design, the different combinations of eHMI types across AVs in Lanes 1 and 2 resulted in nine distinct scenarios. ANOVA was first performed to compare the impacts of allocentric and egocentric communication on cognitive load. Two groups with no eHMI on Lane 2 AVs were selected to eliminate the effects of distractions from vehicles in Lane 2, which isolated the impact of the eHMIs on the Lane 1 AVs. The results, shown in Table 7, indicate that the eHMI type had a significant main effect on the proportion of the pedestrian's gaze on the eHMI ($F(1, 179) = 5.818$, $p < 0.05$, $\eta p^2 = 0.032$). Post hoc comparisons were conducted via an least significant difference (LSD) test (Williams and Abdi, 2010), as shown in Fig. 15(a). When no eHMI was present on the Lane 2 AVs, allocentric communication from the Lane 1 AVs ($M = 0.194$, $SD = 0.143$) led to a greater cognitive load than egocentric communication ($M = 0.144$, $SD = 0.130$). This suggests that allocentric communication may be more difficult for pedestrians to interpret than egocentric communication.

**Table 7.** ANOVA results related to effect of eHMI type on proportion of gaze directed at eHMI on Lane 1 AV when no eHMI is present on Lane 2 AV.

| Variable | Lane 1 eHMI type | Descriptives | | | ANOVA | | | |
| --- | --- | --- | --- | --- | --- | --- | --- | --- |
| | | Sample size | Mean | SD | $df$ | $F$ | $p$ | $\eta p^2$ |
| Gaze proportion for Lane 1 eHMI | 1 | 45 | 0.194 | 0.143 | (1,179) | 5.818 | * | 0.032 |
| | 2 | 45 | 0.144 | 0.130 | | | | |

* p ≤ 0.05, ** p ≤ 0.01, *** p ≤ 0.001

Note: For eHMI type, "1" = allocentric and "2" = egocentric

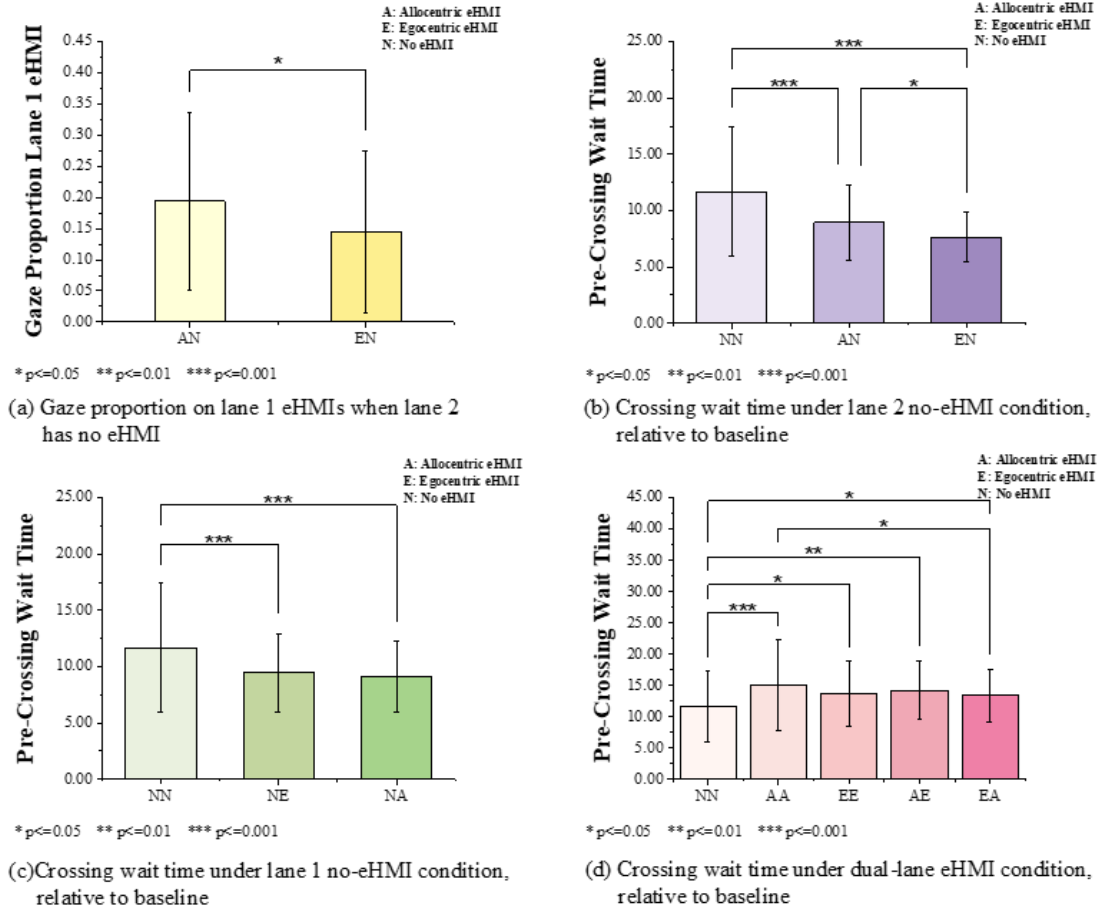

**Fig. 15.** ANOVA results related to cognitive load, with error bars representing the standard error.

ANOVA was also conducted to compare the pedestrian wait times across different scenarios, including the baseline group (no eHMI on AVs in either lane) and two scenarios where only Lane 1 featured an eHMI-equipped AV. The results, presented in Table 8, show that the eHMI type had a significant main effect on the pedestrian wait time before crossing ($F$ (2, 269) = 23.444, $p$ < 0.001, $\eta p^2$ = 0.149). Fig. 15(b) displays the results of post hoc comparisons performed through an LSD test. When no eHMI was present on any Lane 2 AV ($M$ = 11.669, $SD$ = 5.712), the presence of an eHMI on a Lane 1 AV helped pedestrians make crossing decisions faster, which indicates that a single eHMI can reduce cognitive load. In terms of the eHMI type, egocentric communication ($M$ = 7.630, $SD$ = 2.214) resulted in a lower cognitive load than allocentric communication ($M$ = 8.935, $SD$ = 3.376).

**Table 8.** ANOVA results related to effect of eHMI type on pre-crossing wait time for Lane 1 compared with baseline scenario.

| Variable | Descriptives | ANOVA |
| --- | --- | --- |

| | Lane 1 eHMI type | Sample size | Mean | SD | $df$ | $F$ | $p$ | $\eta p^2$ |
|---|---|---|---|---|---|---|---|---|
| Pre-crossing wait time | 0 | 45 | 11.669 | 5.712 | | | | |
| | 1 | 45 | 8.935 | 3.376 | (2,269) | 23.444 | *** | 0.149 |
| | 2 | 45 | 7.630 | 2.214 | | | | |

\* p ≤ 0.05, \*\* p ≤ 0.01, \*\*\* p ≤ 0.001

Note: For eHMI type, "0" = no eHMI, "1" = allocentric, and "2" = egocentric

Next, the baseline group was compared with two scenarios where only Lane 2 AVs featured eHMIs. The results, shown in Table 9, highlight that the eHMI type had a significant main effect on the pedestrian wait time ($F(2, 269) = 9.574$, $p < 0.001$, $\eta p^2 = 0.067$). Fig. 15(c) presents the results of post hoc comparisons performed via an LSD test. When Lane 1 had no eHMI-equipped AV, the presence of an eHMI on a Lane 2 AV reduced the decision-making time for the pedestrians. However, no significant difference was found between egocentric ($M = 9.099$, $SD = 3.149$) and allocentric ($M = 9.452$, $SD = 3.491$) communication, which could be due to the pedestrians focusing more on the AVs in Lane 1 before crossing.

**Table 9.** ANOVA results related to effect of eHMI type on pre-crossing wait time for Lane 2 compared with baseline scenario.

| Variable | Lane 2 eHMI type | Descriptives | | | ANOVA | | | |
|---|---|---|---|---|---|---|---|---|
| | | Sample size | Mean | SD | $df$ | $F$ | $p$ | $\eta p^2$ |
| Pre-crossing wait time | 0 | 45 | 11.669 | 5.712 | | | | |
| | 1 | 45 | 9.452 | 3.491 | (2,269) | 9.574 | *** | 0.067 |
| | 2 | 45 | 9.099 | 3.149 | | | | |

\* p ≤ 0.05, \*\* p ≤ 0.01, \*\*\* p ≤ 0.001

Note: For eHMI type, "0" = no eHMI, "1" = allocentric, and "2" = egocentric

Finally, the baseline group was compared with five scenarios in which both lanes featured eHMI-equipped AVs. Table 10 shows the results, which indicate that the eHMI type had a significant main effect on the pedestrian wait time ($F(4, 449) = 4.566$, $p < 0.01$, $\eta p^2 = 0.039$). The findings from post hoc comparisons conducted through an LSD test are shown in Fig. 15(d). When both lanes included eHMI-equipped AVs, the wait time significantly increased over the baseline scenario, which suggests that multiple eHMIs increase cognitive load. The highest cognitive load was observed when both lanes had AVs with allocentric eHMIs ($M = 15.028$, $SD = 7.209$).

**Table 10.** ANOVA results related to effect of eHMI-equipped AVs in both lanes on pre-crossing wait time compared with baseline scenario.

| Variable | Lane 1 eHMI type | Lane 2 eHMI type | Descriptives | | | ANOVA | | | |
| --- | --- | --- | --- | --- | --- | --- | --- | --- | --- |
| | | | Sample size | Mean | SD | $df$ | $F$ | $p$ | $\eta p^2$ |
| Pre-crossing wait time | 0 | 0 | 45 | 11.669 | 5.712 | (4,449) | 4.566 | ** | 0.039 |
| | 1 | 1 | 45 | 15.028 | 7.209 | | | | |
| | 2 | 2 | 45 | 13.617 | 5.257 | | | | |
| | 1 | 2 | 45 | 14.216 | 4.714 | | | | |
| | 2 | 1 | 45 | 13.402 | 4.238 | | | | |

\* p ≤ 0.05, \*\* p ≤ 0.01, \*\*\* p ≤ 0.001

Note: For eHMI type, "0" = no eHMI, "1" = allocentric, and "2" = egocentric

### 4.3.2 Factors influencing cognitive load

Repeated measures within participants may have introduced intra-individual correlations. GLMMs were used to assess how eHMI combinations affect cognitive load while accounting for within-subject correlations. Cognitive load was assessed through a combination of objective behavioral indicators and subjective self-reported measures. The behavioral indicators included the gaze time for the vehicle in Lane 1 before crossing, the proportion of the gaze directed at the eHMI on a Lane 1 AV, and the pre-crossing wait time. The subjective perception of cognitive load was measured using the NASA-TLX scale. The aforementioned indicators were set as dependent variables in the model. Among these, the gaze time for Lane 1 AVs before crossing (skewness = 2.238, kurtosis = 16.724) and the pre-crossing waiting time (skewness = 1.302, kurtosis = 6.037) represent non-negative and right-skewed continuous variables; therefore, a gamma distribution (Berchialla et al., 2009) with a log-link function was selected for inclusion in the model. The proportion of the gaze on the Lane 1 AV's eHMI constitutes proportional data continuously distributed between 0 and 1; therefore, a beta distribution with a logit-link function was selected for inclusion in the model. The NASA-TLX score (skewness = -0.045, kurtosis = 2.380) followed a normal distribution and was modeled as a Gaussian distribution with a log-link function. To specifically examine how eHMI combinations across different lanes affected cognitive load, the interaction terms between the eHMIs of AVs in both lanes were introduced into the model, as shown in Table 11. The results revealed that the eHMI combination had a significant effect, with cognitive load indicators varying according to changes in the eHMI parameters.

First, regarding the gaze time for the Lane 1 vehicle before crossing, the condition in which vehicles in neither lane displayed an eHMI was selected as the reference scenario. Compared with this baseline, the eHMI type for vehicles in both lanes significantly influenced the duration of the pedestrian's gaze toward the Lane 1 vehicle before crossing. Specifically, relative to the absence of an eHMI on a Lane 1 AV, the presence of an allocentric eHMI ($\beta$ = -0.393, $z$ = -5.20, $p$ < 0.001) and an egocentric eHMI ($\beta$ = -0.447, $z$ = -5.91, $p$ < 0.001) on a Lane 1 AV significantly reduced the gaze time. A similar reduction was observed when Lane 2 AVs displayed either allocentric eHMI ($\beta$ = -0.493, $z$ = -6.50, $p$ < 0.001) or egocentric eHMIs ($\beta$ = -0.598, $z$ = -7.90, $p$ < 0.001). With regard to the interaction effects of eHMI combinations across the two lanes, all four dual-lane configurations significantly increased the duration of the pedestrian's gaze on the Lane 1 vehicle ($\beta$ = 1.118, $z$ = 10.43, $p$ < 0.001; $\beta$ = 1.121, $z$ = 10.47, $p$ < 0.001; $\beta$ = 1.015, $z$ = 9.44, $p$ < 0.001; $\beta$ = 1.258, $z$ = 11.75, $p$ < 0.001). Additionally, the results from the post-experiment questionnaire indicate that age ($\beta$ = -0.051, $z$ = -2.64, $p$ < 0.01) and the level of understanding of eHMIs ($\beta$ = -0.032, $z$ = -2.39, $p$ < 0.05) had a significant negative effect on the gaze time for the Lane 1 AV before crossing, while the effects of the other indicators were not significant.

Second, regarding the proportion of the gaze on the Lane 1 AV's eHMI before crossing, the condition in which the Lane 1 AVs displayed an allocentric eHMI and the Lane 2 AVs had no eHMI was selected as the reference scenario. Relative to this baseline, the presence of an egocentric eHMI on the Lane 1 AV significantly reduced the gaze proportion for the Lane 1 AV's eHMI ($\beta$ = -0.332, $z$ = -3.44, $p$ < 0.001). However, neither the eHMI type on the Lane 2 AV nor the interaction between the eHMIs of the AVs in the two lanes had any significant effect on the gaze proportion for the Lane 1 AV's eHMI before crossing. An analysis of the post-experiment subjective questionnaire data indicates that age ($\beta$ = 0.130, $z$ = 1.99, $p$ < 0.05) had a significant positive effect on the proportion of the gaze toward the eHMIs of the vehicles in Lane 1 before crossing. Moreover, the vehicle yielding strategy ($\beta$ = -0.270, $z$ = -0.090, $p$ < 0.01) had a significant negative effect on the proportion of the gaze toward the eHMI on the Lane 1 AVs before crossing. The effects of the other indicators were not significant.

Regarding the pedestrian pre-crossing wait time, the condition in which AVs in neither lane displayed an eHMI was selected as the reference scenario.

Compared with this baseline, both types of eHMIs on AVs in both lanes significantly reduced the wait time. Specifically, relative to the absence of an eHMI on vehicles in Lane 1, the presence of an allocentric eHMI ($\beta$ = -0.265, $z$ = -5.38, $p$ < 0.001) and an egocentric eHMI ($\beta$ = -0.402, $z$ = -8.18, $p$ < 0.001) significantly reduced the wait time. Similar reductions were observed for allocentric ($\beta$ = -0.240, $z$ = -4.89, $p$ < 0.001) and egocentric eHMIs ($\beta$ = -0.202, $z$ = -4.11, $p$ < 0.001) on Lane 2 vehicles. In contrast, all four combinations of dual-lane eHMI configurations significantly increased the wait time ($\beta$ = 0.771, $z$ = 11.08, $p$ < 0.001; $\beta$ = 0.697, $z$ = 10.04, $p$ < 0.001; $\beta$ = 0.812, $z$ = 11.70, $p$ < 0.001; $\beta$ = 0.774, $z$ = 11.15, $p$ < 0.001). Furthermore, the responses from the subjective questionnaire indicate that age ($\beta$ = 0.051, $z$ = 2.79, $p$ < 0.01) had a significant positive effect on the wait time, whereas the vehicle yielding strategy ($\beta$ = -0.154, $z$ = -6.64, $p$ < 0.001) and crossing adherence level ($\beta$ = -0.019, $z$ = -2.09, $p$ < 0.05) had a significant negative effect. The effects of the other indicators were not significant.

Table 11. GLMM results related to pedestrian cognitive load for different eHMI configurations and individual factors.

| Variable | Gaze time for Lane 1 AV (a) | | | | Gaze proportion for Lane 1 eHMI (b) | | | | Pre-crossing wait time (a) | | | |
|---|---|---|---|---|---|---|---|---|---|---|---|---|
| | $\beta$ | S.E. | z | p | $\beta$ | S.E. | z | p | $\beta$ | S.E. | z | p |
| eHMI_AV1 | | | | | | | | | | | | |
| 1 | -0.393 | 0.076 | -5.20 | *** | — | — | — | — | -0.265 | 0.049 | -5.38 | *** |
| 2 | -0.447 | 0.076 | -5.91 | *** | -0.332 | 0.096 | -3.44 | *** | -0.402 | 0.049 | -8.18 | *** |
| eHMI_AV2 | | | | | | | | | | | | |
| 1 | -0.493 | 0.076 | -6.50 | *** | -0.075 | 0.083 | -0.91 | 0.364 | -0.240 | 0.049 | -4.89 | *** |
| 2 | -0.598 | 0.076 | -7.90 | *** | -0.128 | 0.086 | -1.49 | 0.137 | -0.202 | 0.049 | -4.11 | *** |
| eHMI_AV1*eHMI_AV2 | | | | | | | | | | | | |
| 1 1 | 1.118 | 0.107 | 10.43 | *** | — | — | — | — | 0.771 | 0.070 | 11.08 | *** |
| 1 2 | 1.121 | 0.107 | 10.47 | *** | — | — | — | — | 0.697 | 0.069 | 10.04 | *** |
| 2 1 | 1.015 | 0.107 | 9.44 | *** | 0.059 | 0.140 | 0.42 | 0.675 | 0.812 | 0.069 | 11.70 | *** |
| 2 2 | 1.258 | 0.107 | 11.75 | *** | 0.238 | 0.136 | 1.75 | 0.080 | 0.774 | 0.069 | 11.15 | *** |
| Gender | -0.021 | 0.062 | -0.34 | 0.736 | -0.200 | 0.146 | -1.37 | 0.171 | -0.016 | 0.058 | -0.28 | 0.778 |
| Age | -0.051 | 0.019 | -2.64 | ** | 0.064 | 0.043 | 1.49 | 0.135 | 0.051 | 0.018 | 2.79 | ** |
| Driver's license | 0.056 | 0.114 | -0.49 | 0.623 | 0.306 | 0.283 | 1.08 | 0.281 | 0.024 | 0.106 | 0.23 | 0.818 |
| Driving experience | 0.028 | 0.030 | 0.93 | 0.351 | 0.122 | 0.066 | 1.84 | 0.066 | -0.038 | 0.027 | -1.38 | 0.168 |
| Education | 0.243 | 0.104 | 2.34 | * | -0.196 | 0.238 | -0.82 | 0.410 | -0.150 | 0.098 | -1.54 | 0.124 |
| Practitioner | 0.026 | 0.081 | 0.32 | 0.746 | 0.075 | 0.194 | 0.39 | 0.697 | -0.064 | 0.076 | -0.84 | 0.404 |
| Yielding strategy | -0.041 | 0.036 | -1.14 | 0.256 | -0.224 | 0.056 | -3.97 | *** | -0.154 | 0.023 | -6.64 | *** |
| Understanding of AVs | 0.015 | 0.011 | 1.37 | 0.172 | -0.015 | 0.026 | -0.58 | 0.559 | 0.015 | 0.010 | 1.39 | 0.165 |
| Trust in AVs | 0.011 | 0.008 | 1.36 | 0.174 | 0.030 | 0.019 | 1.58 | 0.113 | 0.006 | 0.008 | 0.74 | 0.460 |
| Risk perception of AVs | -0.015 | 0.009 | -1.59 | 0.111 | -0.039 | 0.022 | -1.77 | 0.077 | -0.004 | 0.009 | -0.40 | 0.689 |
| Understanding of eHMIs | -0.032 | 0.013 | -2.39 | * | 0.008 | 0.031 | 0.27 | 0.788 | -0.005 | 0.012 | -0.36 | 0.716 |
| Trust in eHMIs | -0.001 | 0.011 | -0.10 | 0.918 | 0.008 | 0.024 | 0.34 | 0.737 | 0.016 | 0.010 | 1.61 | 0.107 |
| Crossing focus | 0.001 | 0.009 | 0.09 | 0.927 | 0.042 | 0.025 | 1.67 | 0.096 | 0.017 | 0.010 | 1.66 | 0.096 |
| Crossing adherence | -0.011 | 0.009 | -1.14 | 0.254 | -0.002 | 0.021 | -0.09 | 0.931 | -0.018 | 0.009 | -2.09 | * |
| Attitudes toward others | 0.009 | 0.010 | 0.91 | 0.363 | 0.008 | 0.022 | 0.37 | 0.714 | 0.008 | 0.009 | 0.86 | 0.390 |
| AIC | 4296.253 | | | | -744.242 | | | | 4427.897 | | | |
| BIC | 4418.376 | | | | -645.536 | | | | 4550.020 | | | |
| Log likelihood | -2122.126 | | | | 395.121 | | | | -2187.949 | | | |

\* $p \leq 0.05$, \*\* $p \leq 0.01$, \*\*\* $p \leq 0.001$

Note: The standardized regression coefficients are provided. The reference category for (a) is both lanes without eHMIs, while that for (b) is Lane 1 with allocentric eHMIs. "eHMI_AV1\*eHMI_AV2" refers to the interaction term between eHMI1 and eHMI2. For the eHMI type, "0" = no eHMI, "1" = allocentric, and "2" = egocentric.

Table 12 summarizes the GLMM analysis results based on the NASA-TLX scores. Regarding subjective cognitive load, the condition in which AVs in neither lane displayed an eHMI was selected as the reference scenario. Compared with this baseline, the eHMI types on vehicles in both lanes significantly influenced the subjective cognitive load perceived by the pedestrians. Specifically, relative to the absence of an eHMI on Lane 1 AVs, the presence of an allocentric eHMI ($\beta$ = -0.361, $z$ = -6.51, $p$ < 0.001) and an egocentric eHMI ($\beta$ = -0.442, $z$ = -7.96, $p$ < 0.001) on Lane 1 AVs significantly reduced the subjective workload. Similar reductions were also observed for allocentric eHMIs ($\beta$ = -0.280, $z$ = -5.05, $p$ < 0.001) and egocentric eHMIs ($\beta$ = -0.181, $z$ = -3.28, $p$ < 0.001) on Lane 2 AVs. In contrast, all four combinations of eHMIs across both lanes significantly increased the subjective cognitive load ($\beta$ = 0.775, $z$ = 9.86, $p$ < 0.001; $\beta$ = 0.638, $z$ = 8.13, $p$ < 0.001; $\beta$ = 0.849, $z$ = 10.76, $p$ < 0.001; $\beta$ = 0.769, $z$ = 9.78, $p$ < 0.001).

Additionally, the responses from the post-experiment subjective questionnaires indicate that higher perceived risks from AVs were associated with significantly lower subjective cognitive loads ($\beta$ = -0.034, $z$ = -2.38, $p$ < 0.05) and that the vehicle yielding strategy ($\beta$ = -0.061, $z$ = -2.36, $p$ < 0.05) had a significant negative effect on subjective cognitive load. The effects of the other indicators were not significant.

**Table 12.** GLMM results based on NASA-TLX scores for different eHMI configurations and individual factors.

| Variable | NASA-TLX | | | |
| --- | --- | --- | --- | --- |
| | $\beta$ | S.E. | $z$ | $p$ |
| eHMI_AV1 | | | | |
| 1 | -0.361 | 0.055 | -6.51 | *** |
| 2 | -0.442 | 0.055 | -7.96 | *** |
| eHMI_AV2 | | | | |
| 1 | -0.280 | 0.055 | -5.05 | *** |
| 2 | -0.181 | 0.055 | -3.28 | *** |
| eHMI_AV1*eHMI_AV2 | | | | |
| 1 1 | 0.775 | 0.079 | 9.86 | *** |
| 1 2 | 0.638 | 0.079 | 8.13 | *** |
| 2 1 | 0.849 | 0.079 | 10.76 | *** |
| 2 2 | 0.769 | 0.079 | 9.78 | *** |
| Gender | 0.066 | 0.093 | 0.71 | 0.478 |
| Age | -0.033 | 0.029 | -1.12 | 0.264 |
| Driver's license | 0.291 | 0.170 | 1.71 | 0.088 |
| Driving experience | -0.011 | 0.044 | -0.24 | 0.810 |
| Education | -0.100 | 0.157 | -0.63 | 0.526 |
| Practitioner | -0.026 | 0.122 | -0.21 | 0.833 |
| Yielding strategy | -0.061 | 0.026 | -2.36 | * |
| Understanding of AVs | 0.016 | 0.017 | 0.96 | 0.337 |
| Trust in AVs | -0.009 | 0.013 | -0.70 | 0.484 |
| Risk perception of AVs | -0.034 | 0.014 | -2.38 | * |
| Understanding of eHMIs | 0.022 | 0.020 | 1.07 | 0.283 |
| Trust in eHMIs | -0.030 | 0.016 | -1.84 | 0.066 |
| Crossing focus | -0.003 | 0.017 | -0.19 | 0.849 |
| Crossing adherence | 0.011 | 0.014 | 0.77 | 0.444 |
| Attitudes toward others | -0.004 | 0.014 | -0.27 | 0.784 |
| AIC | | | 6685.132 | |
| BIC | | | 6807.255 | |
| Log likelihood | | | -3316.566 | |

* $p \leq 0.05$, ** $p \leq 0.01$, *** $p \leq 0.001$

Note: For eHMI type, "1" = allocentric and "2" = egocentric

### 4.4 Distraction effect

*4.4.1 Effect of eHMI configurations on distraction*

Heatmaps were used to assess pedestrian gaze toward vehicles in the non-interactive lane. These heatmaps, based on aggregated gaze data, offered insight into how eHMI types influence distraction. As shown in Fig. 16(a), when neither lane had an eHMI-equipped AV, the gaze of the pedestrians toward vehicles in Lane 2 was minimal, which indicates that these vehicles do not distract pedestrians significantly.

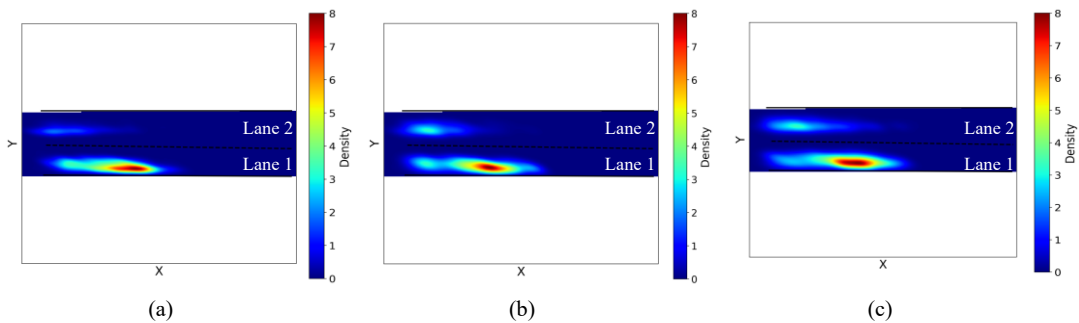

(a)      (b)      (c)

**Fig. 16.** Kernel density heatmap of pedestrian gaze on vehicles in different scenarios: (a) AVs in both lanes without eHMIs; (b) No eHMI + egocentric eHMI; (c) No eHMI + allocentric eHMI.

By comparison, when the Lane 2 AVs were equipped with eHMIs, the density of the pedestrian gaze on these AVs increased significantly, as seen in Fig. 16(b) and 16(c); this suggests that the presence of eHMIs on Lane 2 AVs diverts the attention of pedestrians. Additionally, the distraction effect of allocentric eHMIs is observed to be stronger than that of egocentric eHMIs.

ANOVA was first conducted to examine how different eHMI types on the Lane 2 AVs affected the distraction behavior of the participants, considering the Lane 1 AVs to be equipped with the same type of eHMI. The first comparison focused on the three groups in which the Lane 1 AVs had no eHMI. The results, shown in Table 13, indicate that different eHMI types on Lane 2 AVs had a significant main effect on the proportion of the pedestrian gaze directed toward vehicles in Lane 2 ($F(2, 269) = 23.994$, $p < 0.001$, $\eta p^2 = 0.152$). Post hoc comparisons performed through an LSD test (Fig. 17(a)) revealed that when no eHMI was present on any Lane 1 AV, allocentric communication from Lane 2 AVs ($M = 0.339$, $SD = 0.166$) attracted a higher proportion of the pedestrian gaze than egocentric communication ($M = 0.266$, $SD = 0.197$).

**Table 13.** ANOVA results related to effect of eHMI type on gaze proportion toward Lane 2 vehicles before crossing, relative to the baseline, when no eHMI was present on Lane 1 vehicles.

| Variable | Lane 2 eHMI type | Descriptives | | | ANOVA | | | |
| --- | --- | --- | --- | --- | --- | --- | --- | --- |
| | | Sample size | Mean | SD | $df$ | $F$ | $p$ | $\eta p^2$ |
| Gaze proportion for Lane 2 AV | 0 | 45 | 0.170 | 0.120 | (2,269) | 23.994 | *** | 0.152 |
| | 1 | 45 | 0.339 | 0.166 | | | | |
| | 2 | 45 | 0.266 | 0.197 | | | | |

\* $p \leq 0.05$, ** $p \leq 0.01$, *** $p \leq 0.001$

Note: For eHMI type, "0" = no eHMI, "1" = allocentric, and "2" = egocentric

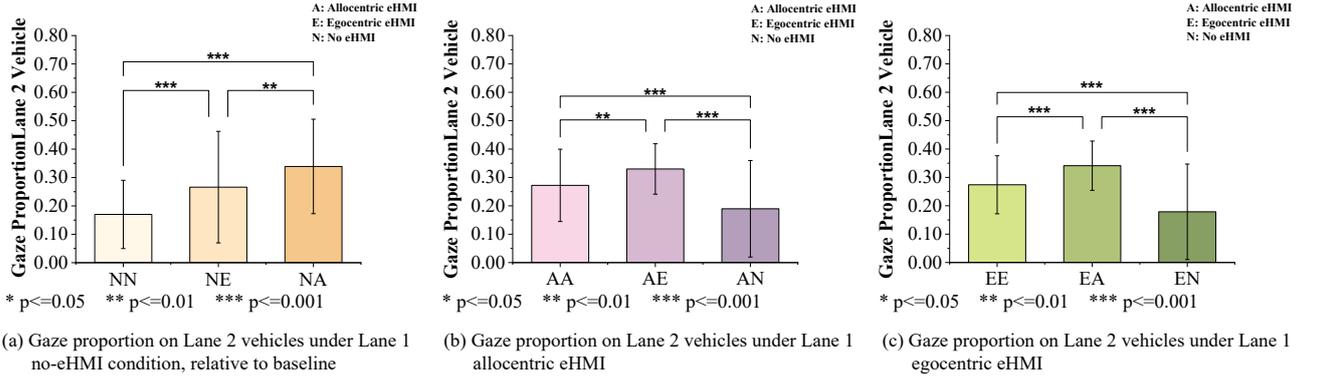

(a) Gaze proportion on Lane 2 vehicles under Lane 1 no-eHMI condition, relative to baseline

(b) Gaze proportion on Lane 2 vehicles under Lane 1 allocentric eHMI

(c) Gaze proportion on Lane 2 vehicles under Lane 1 egocentric eHMI

**Fig. 17.** ANOVA results related to pedestrian distraction, with error bars representing the standard error.

Next, a comparison was conducted between the three groups in which the Lane 1 AVs consistently featured allocentric eHMIs. The results, presented in Table 14, indicate that different eHMI types on Lane 2 AVs had a significant main effect on the proportion of the pedestrian gaze toward Lane 2 AVs ($F(2, 269) = 25.319$, $p < 0.001$, $\eta p^2 = 0.159$). Post hoc comparisons via an LSD test (Fig. 17(b)) revealed that when the Lane 1 AVs consistently displayed allocentric communication, the presence of an eHMI on the Lane 2 AVs caused greater pedestrian distraction than the absence of an eHMI on the Lane 2 AVs ($M = 0.190$, $SD = 0.170$). Among the eHMI types on the Lane 2 AVs, egocentric communication ($M = 0.330$, $SD = 0.089$) attracted a higher proportion of the pedestrian gaze than allocentric communication ($M = 0.272$, $SD = 0.127$).

**Table 14.** ANOVA results related to effect of eHMI type on gaze proportion for Lane 2 vehicles before crossing, relative to the baseline, when Lane 1 vehicles feature allocentric eHMIs.

| Variable | Lane 2 eHMI type | Descriptives | | | ANOVA | | | |
|---|---|---|---|---|---|---|---|---|
| | | Sample size | Mean | SD | df | F | p | $\eta p^2$ |
| Gaze proportion for Lane 2 AV | 0 | 45 | 0.190 | 0.170 | (2,269) | 25.319 | *** | 0.159 |
| | 1 | 45 | 0.272 | 0.127 | | | | |
| | 2 | 45 | 0.330 | 0.089 | | | | |

* p ≤ 0.05, ** p ≤ 0.01, *** p ≤ 0.001

Note: For eHMI, "0" = no eHMI, "1" = allocentric, and "2" = egocentric

Finally, a comparison was conducted between the three groups in which the Lane 1 AVs consistently featured egocentric eHMIs. The results, shown in Table 15, reveal that different eHMI types on the Lane 2 AVs had a significant main effect on the proportion of the pedestrian gaze toward Lane 2 AVs ($F(2,$

269) = 38.789, $p < 0.001$, $\eta p^2 = 0.225$). Post hoc comparisons conducted through an LSD test (Fig. 17(c)) demonstrated that when the Lane 1 AVs consistently displayed egocentric communication, the presence of an eHMI on the Lane 2 AVs caused greater pedestrian distraction than the absence of an eHMI on the Lane 2 AVs ($M = 0.179$, $SD = 0.168$). Among the eHMI types on the Lane 2 AVs, allocentric communication ($M = 0.341$, $SD = 0.087$) attracted a higher proportion of the pedestrian gaze than egocentric communication ($M = 0.274$, $SD = 0.102$).

**Table 15.** ANOVA results related to effect of eHMI type on gaze proportion for Lane 2 vehicles before crossing, relative to the baseline, when Lane 1 featured AVs with egocentric eHMIs.

| Variable | Lane 2 eHMI type | Descriptives | | | ANOVA | | | |
| --- | --- | --- | --- | --- | --- | --- | --- | --- |
| | | Sample size | Mean | SD | $df$ | $F$ | $p$ | $\eta p^2$ |
| Gaze proportion for Lane 2 AV | 0 | 45 | 0.179 | 0.168 | | | | |
| | 1 | 45 | 0.341 | 0.087 | (2,269) | 38.789 | *** | 0.225 |
| | 2 | 45 | 0.274 | 0.102 | | | | |

\* $p \leq 0.05$, \*\* $p \leq 0.01$, \*\*\* $p \leq 0.001$

Note: For eHMI type, "0" = no eHMI, "1" = allocentric, and "2" = egocentric

*4.4.2 Factors influencing pedestrian distraction*

Distraction was measured by gaze proportions on Lane 2 vehicles and their eHMIs before crossing. As both variables represent continuous proportional data bounded between 0 and 1, a beta distribution with a logit-link function was selected for the model (Verkuilen and Smithson, 2012). Interaction terms between Lane 1 and Lane 2 eHMI types were included to assess cross-lane distraction effects. The results, presented in Table 16, reveal that differences in the configuration of eHMI types caused significant variations in the distraction behavior of the participants.

First, the proportion of the gaze toward vehicles in Lane 2 before crossing was examined. The condition in which neither lane featured an eHMI-equipped AV was selected as the reference scenario. Relative to this baseline, the presence of eHMIs on the Lane 2 AVs significantly influenced the visual attention of the pedestrians to these vehicles before crossing. Specifically, relative to the absence of an eHMI, the presence of an allocentric eHMI ($\beta = 0.685$, $z = 7.47$, $p$

< 0.001) and an egocentric eHMI ($\beta$ = 0.478, $z$ = 4.95, $p$ < 0.001) on the Lane 2 AVs significantly increased the gaze proportion. With regard to the interaction effects, when AVs in both lanes displayed allocentric eHMIs, the gaze proportion for the Lane 2 vehicles was significantly lower ($\beta$ = -0.353, $z$ = -2.75, $p$ < 0.001). Additionally, the responses of the participants from the post-experiment subjective questionnaire show that age ($\beta$ = 0.035, $z$ = 2.02, $p$ < 0.05) and trust in AVs ($\beta$ = 0.023, $z$ = 3.06, $p$ < 0.01) had significant positive effects on the proportion of the gaze toward vehicles in Lane 2 before crossing. The effects of the other indicators were not significant.

Next, the proportion of the gaze toward the eHMIs on Lane 2 AVs before crossing was examined. The condition in which the Lane 1 AVs had no eHMI and the Lane 2 AVs featured an allocentric eHMI was selected as the reference scenario. Compared with this baseline, the presence of an egocentric eHMI on the Lane 2 AVs significantly reduced the proportion of the gaze toward the eHMIs on the Lane 2 AVs ($\beta$ = -0.653, $z$ = -3.20, $p$ < 0.001). The eHMI type on the Lane 1 AVs also significantly influenced the gaze behavior for the eHMIs on the Lane 2 AVs. Specifically, the presence of an allocentric eHMI ($\beta$ = -0.473, $z$ = -2.63, $p$ < 0.01) and an egocentric eHMI ($\beta$ = -0.444, $z$ = -2.49, $p$ < 0.05) on the Lane 1 AVs both significantly reduced the proportion of the gaze directed at the eHMIs on the Lane 2 AVs. Regarding the interaction effects, the condition where the Lane 1 AVs had no eHMI and the Lane 2 AVs had an allocentric eHMI was selected as the reference scenario. Relative to this baseline, the configuration in which both lanes had AVs displaying egocentric eHMIs significantly reduced the proportion of the gaze directed toward the eHMIs on the Lane 2 AVs ($\beta$ = -1.117, $z$ = -4.11, $p$ < 0.001). Regarding the subjective questionnaire responses, pedestrians with more driving experience ($\beta$ = 0.221, $z$ = 2.45, $p$ < 0.05) paid greater attention to the Lane 2 AV's eHMI before crossing, while no other variables related to individual differences showed any significant effect.

Table 16. GLMM results related to pedestrian distraction for different eHMI configurations and individual factors.

| Variable | Gaze proportion for Lane 2 AV (a) | | | | Gaze proportion for Lane 2 eHMI (b) | | | |
| --- | --- | --- | --- | --- | --- | --- | --- | --- |
| | $\beta$ | S.E. | $z$ | $p$ | $\beta$ | S.E. | $z$ | $p$ |
| eHMI_AV1 | | | | | | | | |
| 1 | 0.117 | 0.110 | 1.07 | 0.284 | -0.473 | 0.180 | -2.63 | ** |
| 2 | 0.089 | 0.111 | 0.80 | 0.424 | -0.444 | 0.178 | -2.49 | * |
| eHMI_AV2 | | | | | | | | |
| 1 | 0.685 | 0.091 | 7.47 | *** | — | — | — | — |
| 2 | 0.478 | 0.096 | 4.95 | *** | -0.653 | 0.204 | -3.20 | *** |
| eHMI_AV1* eHMI_AV2 | | | | | | | | |
| 1 1 | 0.353 | 0.128 | 2.75 | ** | — | — | — | — |
| 1 2 | 0.031 | 0.129 | 0.24 | 0.809 | 0.445 | 0.317 | 1.40 | 0.160 |
| 2 1 | -0.117 | 0.126 | -0.92 | 0.357 | — | — | — | — |
| 2 2 | -0.118 | 0.133 | -0.88 | 0.357 | -1.117 | 0.272 | -4.11 | *** |
| Gender | -0.030 | 0.056 | -0.53 | 0.594 | 0.054 | 0.174 | 0.31 | 0.756 |
| Age | 0.035 | 0.017 | 2.02 | * | 0.045 | 0.058 | -0.78 | 0.433 |
| Driver's license | 0.117 | 0.109 | 1.07 | 0.283 | -0.205 | 0.351 | -0.58 | 0.559 |
| Driving experience | 0.013 | 0.027 | 0.48 | 0.628 | 0.221 | 0.090 | 2.45 | * |
| Education | -0.171 | 0.094 | -1.81 | 0.070 | 0.296 | 0.319 | 0.93 | 0.355 |
| Practitioner | 0.054 | 0.072 | 0.75 | 0.451 | 0.210 | 0.248 | 0.85 | 0.397 |
| Yielding strategy | -0.027 | 0.035 | -0.79 | 0.430 | -0.084 | 0.106 | -0.79 | 0.429 |
| Understanding of AVs | -0.007 | 0.010 | -0.71 | 0.480 | -0.049 | 0.033 | -1.47 | 0.141 |
| Trust in AVs | 0.023 | 0.007 | 3.06 | ** | 0.011 | 0.024 | 0.47 | 0.639 |
| Risk perception of AVs | -0.004 | 0.009 | -0.45 | 0.656 | 0.009 | 0.025 | 0.36 | 0.717 |
| Understanding of eHMIs | -0.003 | 0.012 | -0.24 | 0.812 | 0.003 | 0.038 | 0.08 | 0.936 |
| Trust in eHMIs | -0.002 | 0.010 | -0.18 | 0.859 | 0.004 | 0.029 | 0.12 | 0.902 |
| Crossing focus | -0.018 | 0.010 | -1.76 | 0.079 | 0.049 | 0.032 | 1.57 | 0.118 |
| Crossing adherence | -0.001 | 0.009 | -0.06 | 0.950 | 0.004 | 0.025 | 0.16 | 0.873 |
| Attitudes toward others | 0.003 | 0.009 | 0.40 | 0.691 | 0.010 | 0.025 | 0.38 | 0.706 |
| AIC | | | -876.001 | | | | -371.254 | |
| BIC | | | -753.879 | | | | -272.55 | |
| Log likelihood | | | 464.001 | | | | 208.63 | |

\* p ≤ 0.05, \*\* p ≤ 0.01, \*\*\* p ≤ 0.001

Note: The standardized regression coefficients are provided. The reference category for (a) is both lanes without eHMIs, while that for (b) is Lane 2 with allocentric eHMIs. For the eHMI type, "1" = allocentric and "2" = egocentric.

## 4.5 Misleading effect

*4.5.1 Group differences in misleading behavior*

Because the dependent variable was a binary outcome expressed as a proportion, a Z-test for proportions was conducted to determine the fraction of pedestrians who were actually misled under different potentially misleading eHMI combinations. This test evaluates whether the difference between two proportions is statistically significant. Table 17 shows the descriptive results, while Fig. 18 presents pairwise comparisons corrected through Bonferroni adjustment.

Under Misleading Behavior 1, the eHMI type on the vehicle in Lane 1 had a significant effect on the proportion of pedestrians who were actually misled. Specifically, when no eHMI was present on the Lane 2 AV, an egocentric eHMI on the Lane 1 AV resulted in a significantly higher proportion of misled pedestrians (*Frequency* = 42.2%) than when the Lane 1 AV had an allocentric eHMI (*Frequency* = 15.6%). Similarly, when the Lane 2 AV had an allocentric eHMI, the proportion of pedestrians misled by an allocentric eHMI on the Lane 1 AV was significantly higher (*Frequency* = 28.9%) than the proportion of pedestrians misled by an egocentric eHMI on the Lane 1 AV (*Frequency* = 4.4%). This pattern was also observed when the Lane 2 AV had an egocentric eHMI.

Under Misleading Behavior 2, the eHMI type on the Lane 2 vehicle significantly affected the proportion of pedestrians who were actually misled. Specifically, when the Lane 1 AV had no eHMI, an egocentric eHMI on the Lane 2 AV resulted in a significantly higher proportion of misled pedestrians (*Frequency* = 53.3%) compared with an allocentric eHMI (*Frequency* = 11.1%). A similar pattern was observed when the Lane 1 AV had an allocentric eHMI: an egocentric eHMI on the Lane 2 AV (*Frequency* = 46.7%) misled a significantly higher proportion of pedestrians than an allocentric eHMI did (*Frequency* = 11.1%).

**Table 17.** Descriptive results related to proportion of pedestrians actually misled across different scenarios by two potentially misleading behaviors.

| Variable | eHMI combination | Descriptives | | |
|---|---|---|---|---|
| | | Sample size | Count (n = 1) | Frequency |
| Misleading Behavior 1 | 1 1 | 45 | 13 | 28.9% |
| | 2 2 | 45 | 2 | 4.4% |
| | 1 2 | 45 | 14 | 31.1% |

| | 2 1 | 45 | 2 | 4.4% |
|---|---|---|---|---|
| | 1 0 | 45 | 7 | 15.6% |
| | 2 0 | 45 | 19 | 42.2% |
| | 1 1 | 45 | 3 | 6.7% |
| | 2 2 | 45 | 21 | 46.7% |
| Misleading Behavior 2 | 1 2 | 45 | 4 | 8.9% |
| | 2 1 | 45 | 5 | 11.1% |
| | 0 2 | 45 | 24 | 53.3% |
| | 0 1 | 45 | 5 | 11.1% |

Note: For eHMI type, "0" = no eHMI, "1" = allocentric, and "2" = egocentric

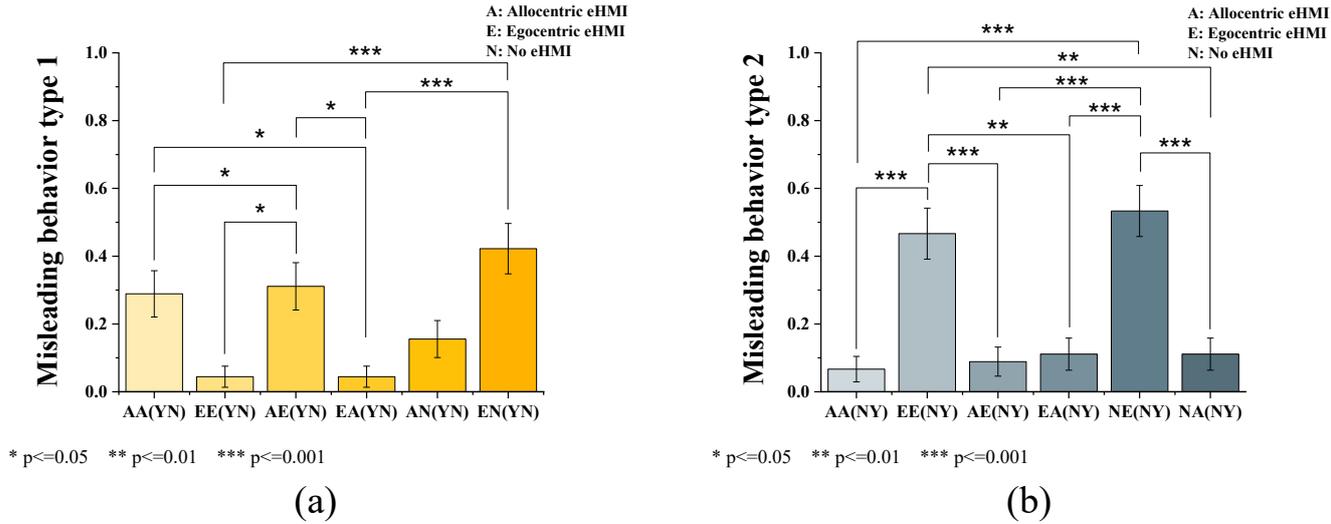

**Fig. 18.** Z-test-for-proportions results related to proportion of pedestrians actually misled by two potentially misleading behaviors: (a) Misleading Behavior 1; (b) Misleading Behavior 2.

*4.5.2 Factors influencing misleading effects*

Because the indicators of misleading effects are binary variables, the logit function, which is the most commonly used link function for binary regression analysis, was chosen as the link function for the GLMM. Misleading Behaviors 1 and 2 were set as dependent variables, while the eHMI types on vehicles in both lanes, their interaction term, and the questionnaire indicators were configured as independent variables. The results are presented in Table 18.

Under Misleading Behavior 1, compared with an allocentric eHMI on the Lane 1 AVs, an egocentric eHMI significantly increased the proportion of pedestrians who were actually misled ($\beta = 1.578$, $z = 2.89$, $p < 0.01$). Specifically, regardless of the type of eHMI on the Lane 2 AVs, the interaction effects between the eHMI types on the AVs in the two lanes significantly reduced the occurrence of Misleading Behavior 1 ($\beta = -3.934$, $z = -3.96$, $p$

< 0.001; $\beta$ = –4.056, $z$ = –4.09, $p$ < 0.001). After including the interaction terms between the eHMI types on the Lane 1 and Lane 2 AVs in the regression model, the marginal effects for each interaction combination were estimated, and pairwise comparisons were conducted. Furthermore, Bonferroni correction was applied to adjust for multiple comparisons. The results again showed that when no eHMI was present on the Lane 2 AVs, an egocentric eHMI on the Lane 1 AVs significantly increased the likelihood of pedestrians being actually misled, compared with an allocentric eHMI (*Contrast* = 0.268, $z$ = 3.14, $p$ < 0.05). In contrast, when the Lane 2 vehicles were equipped with eHMI, whether allocentric (*Contrast* = –0.244, $z$ = 3.49, $p$ < 0.01) or egocentric (Contrast = –0.266, z = 3.75, p < 0.01), the presence of an eHMI on the Lane 2 vehicles mitigated the likelihood of pedestrians being actually misled to some extent. Additionally, a significantly lower proportion of male pedestrians were misled ($\beta$ = –0.949, $z$ = –2.00, $p$ < 0.05), while older pedestrians exhibited a higher tendency to be misled ($\beta$ = 0.34, $z$ = 2.73, $p$ < 0.05). The other individual-level variables did not show any significant effects.

Under Misleading Behavior 2, compared with an allocentric eHMI on the Lane 2 AVs, an egocentric eHMI significantly increased the proportion of pedestrians who were actually misled ($\beta$ = 2.375, $z$ = 4.08, $p$ < 0.001). In contrast, an allocentric eHMI on the Lane 1 AVs significantly reduced the occurrence of Misleading Behavior 2. Specifically, when the Lane 1 AVs were equipped with an allocentric eHMI, the interaction effect between the eHMI types on the AVs in the two lanes significantly lowered the likelihood of pedestrians being misled ($\beta$ = –2.054, $z$ = –2.07, $p$ < 0.05). Pairwise comparisons were conducted again using the marginal-effects-analysis approach. The results indicated that when the Lane 1 vehicles were all equipped with eHMIs, an egocentric eHMI on the Lane 2 vehicles significantly increased the likelihood of pedestrians being actually misled, compared with an allocentric eHMI (*Contrast* = 0.422, $z$ = 4.99, $p$ < 0.001). Similarly, when the Lane 1 vehicles were all equipped with egocentric eHMIs, an egocentric eHMI on the Lane 2 vehicles also led to a significantly higher likelihood of pedestrians being misled, compared with an allocentric eHMI (*Contrast* = 0.356, $z$ = 4.20, $p$ < 0.001). However, this difference was not statistically significant when the Lane

AVs had allocentric eHMIs. The other variables did not show any significant effects.

Table 18. GLMM results related to misleading effects for different eHMI configurations and individual factors.

| Variable | Misleading Behavior 1 (a) | | | | Misleading Behavior 2 (b) | | | |
|---|---|---|---|---|---|---|---|---|
| | $\beta$ | S.E. | $z$ | $p$ | $\beta$ | S.E. | $z$ | $p$ |
| eHMI_AV1 | | | | | | | | |
| 1 | — | — | — | — | -0.578 | 0.775 | -0.75 | 0.456 |
| 2 | 1.578 | 0.547 | 2.89 | ** | 0.001 | 0.684 | 0.01 | 0.954 |
| eHMI_AV2 | | | | | | | | |
| 1 | 0.890 | 0.559 | 1.59 | 0.111 | — | — | — | — |
| 2 | 1.011 | 0.555 | 1.82 | 0.068 | 2.375 | 0.583 | 4.08 | *** |
| eHMI_AV1* eHMI_AV2 | | | | | | | | |
| 1 1 | — | — | — | — | — | — | — | — |
| 1 2 | — | — | — | — | -2.054 | 0.994 | -2.07 | * |
| 2 1 | -3.934 | 0.993 | -3.96 | *** | — | — | — | — |
| 2 2 | -4.056 | 0.992 | -4.09 | *** | -0.293 | 0.815 | -0.36 | 0.720 |
| Gender | -0.949 | 0.475 | -2.00 | * | 0.313 | 0.435 | 0.72 | 0.472 |
| Age | 0.302 | 0.129 | 2.34 | * | -0.149 | 0.144 | -1.03 | 0.303 |
| Driver's license | 0.216 | 0.805 | 0.27 | 0.788 | -1.103 | 0.769 | -1.43 | 0.152 |
| Driving experience | -0.133 | 0.199 | -0.67 | 0.504 | 0.104 | 0.229 | 0.46 | 0.649 |
| Education | -1.288 | 0.683 | -1.89 | 0.059 | -0.072 | 0.747 | -0.10 | 0.924 |
| Practitioner | 0.783 | 0.615 | 1.27 | 0.203 | -0.493 | 0.564 | -0.87 | 0.382 |
| Understanding of AVs | -0.083 | 0.081 | -1.02 | 0.307 | -0.030 | 0.077 | -0.39 | 0.696 |
| Trust in AVs | 0.074 | 0.057 | 1.30 | 0.195 | -0.029 | 0.059 | -0.48 | 0.630 |
| Risk perception of AVs | -0.048 | 0.064 | -0.75 | 0.452 | 0.082 | 0.067 | 1.24 | 0.217 |
| Understanding of eHMIs | 0.073 | 0.095 | 0.77 | 0.440 | 0.131 | 0.097 | 1.34 | 0.179 |
| Trust in eHMIs | 0.098 | 0.079 | 1.25 | 0.212 | -0.051 | 0.077 | -0.66 | 0.507 |
| Crossing focus | 0.088 | 0.077 | 1.15 | 0.250 | -0.115 | 0.078 | -1.48 | 0.139 |
| Crossing adherence | -0.045 | 0.066 | -0.68 | 0.499 | 0.115 | 0.072 | 1.60 | 0.110 |
| Attitudes toward others | 0.017 | 0.067 | 0.25 | 0.802 | -0.115 | 0.068 | -1.70 | 0.089 |
| AIC | | | 284.770 | | | | 262.442 | |
| BIC | | | 349.542 | | | | 334.411 | |
| Log likelihood | | | -124.385 | | | | -111.221 | |

\* p ≤ 0.05, \*\* p ≤ 0.01, \*\*\* p ≤ 0.001

Note: The standardized regression coefficients are provided. The reference category for (a) is Lane 1 with allocentric eHMIs, while that for (b) is Lane 2 with allocentric eHMIs. For the eHMI type, "0" = no eHMI, "1" = allocentric, and "2" = egocentric

## 4.6 Crossing risk

*4.6.1 Group differences in crossing risk*

ANOVA was used to analyze the effect of different eHMI combinations on the pedestrian collision risk indicator, PET, for Lane 1. First, the baseline group, in which no eHMI was present on vehicles in either lane, was compared with combinations where only one lane had vehicles equipped with eHMIs. The results, shown in Table 19, indicate that relative to the scenario without eHMIs on vehicles in either lane, all single-lane eHMI combinations significantly reduced the PET value for pedestrians crossing Lane 1 ($F(4, 449) = 5.284$, $p < 0.001$, $\eta p^2 = 0.045$). Post hoc comparisons performed through an LSD test (Fig. 19(a)) revealed that the presence of eHMIs on Lane 1 AVs ($M = 1.675$, $SD = 1.317$; $M = 1.817$, $SD = 1.574$) was associated with a slightly higher collision risk than the presence of eHMIs on Lane 2 AVs ($M = 2.006$, $SD = 1.629$; $M = 1.932$, $SD = 1.592$).

Next, the baseline group was compared with combinations in which AVs in both lanes were equipped with eHMIs. The results, shown in Table 19, indicate that the presence of eHMIs on vehicles in both lanes also significantly reduced the PET value for pedestrians crossing Lane 1 ($F(4, 449) = 18.910$, $p < 0.001$, $\eta p^2 = 0.145$). Post hoc comparisons via an LSD test (Fig. 19(b)) revealed that mixed combinations, such as allocentric communication from Lane 1 vehicles and egocentric communication from Lane 2 vehicles ($M = 1.208$, $SD = 0.815$), led to a higher collision risk than scenarios where both lanes had vehicles featuring allocentric eHMIs ($M = 1.611$, $SD = 1.254$). Similarly, egocentric communication from Lane 1 AVs combined with allocentric communication from Lane 2 AVs ($M = 1.422$, $SD = 1.017$) resulted in a higher collision risk compared with vehicles in both lanes using egocentric communication ($M = 1.907$, $SD = 1.415$).

**Table 19.** ANOVA results related to effect of eHMI type on PET for Lane 1, relative to the baseline.

| Variable | eHMI combination | Descriptives | | | ANOVA | | | |
|---|---|---|---|---|---|---|---|---|
| | | Sample size | Mean | SD | $df$ | $F$ | $p$ | $\eta p^2$ |
| PET1 | 0 0 | 45 | 2.631 | 1.412 | (4,449) | 5.284 | *** | 0.045 |
| | 0 1 | 45 | 1.982 | 1.592 | | | | |
| | 0 2 | 45 | 2.006 | 1.629 | | | | |
| | 1 0 | 45 | 1.675 | 1.317 | | | | |
|      | 2 0 | 45 | 1.817 | 1.574 |         |        |     |       |
|      | 0 0 | 45 | 2.631 | 1.412 |         |        |     |       |
|      | 1 1 | 45 | 1.611 | 1.254 |         |        |     |       |
| PET2 | 2 2 | 45 | 1.907 | 1.415 | (4,449) | 18.910 | *** | 0.145 |
|      | 1 2 | 45 | 1.208 | 0.815 |         |        |     |       |
|      | 2 1 | 45 | 1.422 | 1.017 |         |        |     |       |

\* p ≤ 0.05, \*\* p ≤ 0.01, \*\*\* p ≤ 0.001

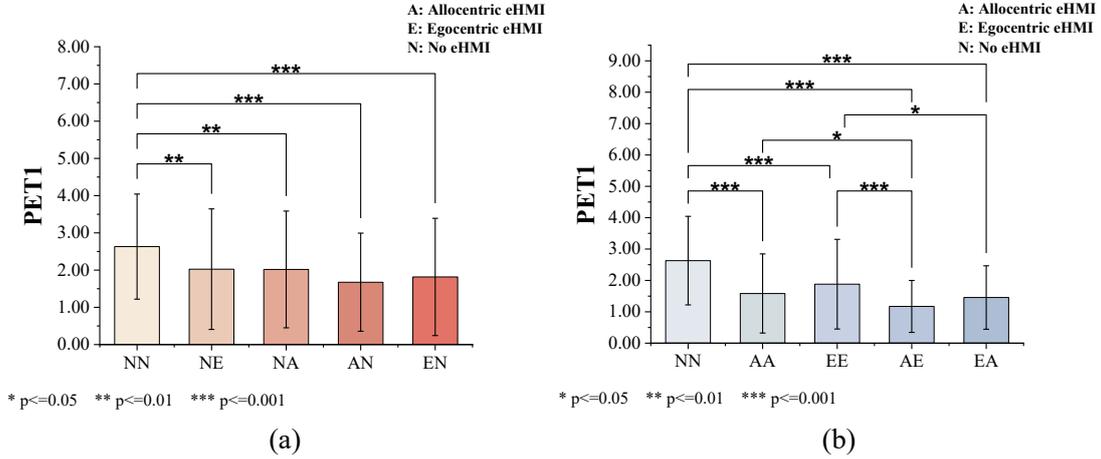

**Fig. 19.** ANOVA results related to PET for pedestrians in Lane 1: (a) Comparison between single-lane eHMI and baseline; (b) Comparison between multi-lane eHMI and baseline (error bars represent the standard error).

*4.6.2 Factors influencing pedestrian crossing risk*

Because the conflict indicators are binary variables (designated as "1" for serious conflicts and "0" for non-serious conflicts, based on the PET thresholds), the logit function was chosen again as the link function for the GLMM. The indicators of cognitive load, distraction, and misleading behavior mentioned earlier were introduced into the model as independent variables, along with the subjective questionnaire indicators. The results are presented in Table 20.

Regarding conflicts in Lane 1, the type of eHMI had a significant main effect. Specifically, the presence of a single eHMI, whether on vehicles in Lane 1 or 2, significantly increased the crossing risk in Lane 1. Specifically, allocentric eHMIs on Lane 1 AVs ($\beta = 1.962$, $z = 4.59$, $p < 0.001$), egocentric eHMIs on Lane 1 AVs ($\beta = 1.466$, $z = 3.50$, $p < 0.001$), and allocentric eHMIs on Lane 2 AVs ($\beta = 1.131$, $z = -2.75$, $p < 0.01$) all increased the risk significantly. However, certain interaction effects between eHMI types significantly alleviated this risk, specifically when both lanes had AVs equipped with allocentric eHMIs ($\beta = -1.748$, $z = -3.02$, $p < 0.01$), or when Lanes 1 and 2 had AVs featuring egocentric and allocentric eHMIs, respectively ($\beta = -1.494$, $z = -2.61$, $p < 0.01$).

Additionally, misleading effects from AVs in Lane 2 ($\beta = 1.668$, $z = 3.64$, $p < 0.001$) significantly increased the crossing risk in Lane 1. In terms of cognitive load, a shorter gaze time for Lane 1 vehicles before crossing ($\beta = -0.067$, $z = -2.49$, $p < 0.05$) was linked to a higher risk. Regarding distraction, a higher gaze proportion for Lane 2 vehicles ($\beta = 1.729$, $z = 2.42$, $p < 0.05$) was significantly associated with an elevated crossing risk in Lane 1. The subjective questionnaire responses showed that having professional backgrounds ($\beta = 1.442$, $z = 3.48$, $p < 0.001$) and greater trust in AVs ($\beta = 0.118$, $z = 2.77$, $p < 0.01$) or eHMIs ($\beta = 0.146$, $z = 2.66$, $p < 0.01$) was associated with an increased crossing risk. In contrast, participants with greater knowledge of eHMI systems were exposed to significantly lower crossing risks ($\beta = -0.207$, $z = -3.08$, $p < 0.01$).

The analysis of conflicts in Lane 2 revealed that the eHMI type had a significant main effect. The presence of a single type of eHMI on Lane 2 AVs significantly increased the crossing risk, particularly when the vehicle was equipped with an allocentric eHMI ($\beta = 1.475$, $z = 3.88$, $p < 0.001$) or egocentric eHMI ($\beta = 0.904$, $z = 2.55$, $p < 0.05$).

Additionally, misleading effects from Lane 1 AVs ($\beta = 0.877$, $z = 2.27$, $p < 0.05$) significantly increased the crossing risk in Lane 2. Regarding distraction-related indicators, a higher proportion of the gaze being directed toward the eHMI on the Lane 2 AV before crossing ($\beta = -1.349$, $z = 2.18$, $p < 0.05$) was significantly associated with reduced risk, which suggests that paying visual attention to eHMI content leads to safer crossing decisions. None of the subjective questionnaire variables showed any statistically significant effect on the crossing risk in Lane 2.

Table 20. GLMM results related to serious AV–pedestrian conflicts in Lanes 1 and 2.

| Variable | Conflict in Lane 1 | | | | Conflict in Lane 2 | | | |
|---|---|---|---|---|---|---|---|---|
| | $\beta$ | S.E. | z | p | $\beta$ | S.E. | z | p |
| eHMI_AV1 | | | | | | | | |
| 1 | 1.961 | 0.427 | 4.59 | *** | 0.408 | 0.383 | 1.07 | 0.287 |
| 2 | 1.466 | 0.419 | 3.50 | *** | 0.121 | 0.388 | 0.31 | 0.756 |
| eHMI_AV2 | | | | | | | | |
| 1 | 1.131 | 0.412 | 2.75 | ** | 1.475 | 0.380 | 3.88 | *** |
| 2 | 0.524 | 0.415 | 1.26 | 0.206 | 0.904 | 0.355 | 2.55 | * |
| eHMI_AV1* eHMI_AV2 | | | | | | | | |
| 1 1 | -1.748 | 0.578 | -3.02 | ** | -0.539 | 0.546 | -0.99 | 0.323 |
| 1 2 | -0.947 | 0.577 | -1.64 | 0.101 | -0.371 | 0.526 | -0.71 | 0.480 |
| 2 1 | -1.494 | 0.573 | -2.61 | ** | -0.695 | 0.538 | -1.29 | 0.196 |
| 2 2 | -0.987 | 0.602 | -1.64 | 0.101 | -0.150 | 0.546 | -0.27 | 0.784 |
| Misleading Behavior 1 | — | — | — | — | 0.877 | 0.386 | 2.27 | * |
| Misleading Behavior 2 | 1.668 | 0.458 | 3.64 | *** | — | — | — | — |
| Gaze time for Lane 1 AV | -0.067 | 0.027 | -2.49 | * | 0.009 | 0.025 | 0.35 | 0.728 |
| Gaze proportion for Lane 1 eHMI | -1.173 | 0.927 | -1.27 | 0.206 | -0.209 | 0.853 | -0.24 | 0.807 |
| Pre-crossing wait time | 0.016 | 0.023 | 0.70 | 0.485 | -0.001 | 0.022 | -0.05 | 0.957 |
| NASA-TLX | 0.008 | 0.007 | 1.29 | 0.198 | -0.005 | 0.006 | -0.86 | 0.390 |
| Gaze proportion for Lane 2 AV | 1.729 | 0.714 | 2.42 | * | -0.625 | 0.629 | -0.99 | 0.321 |
| Gaze proportion for Lane 2 eHMI | 1.219 | 0.656 | 1.86 | 0.063 | -1.349 | 0.619 | -2.18 | * |
| Gender | 0.077 | 0.309 | 0.25 | 0.803 | -0.195 | 0.293 | -0.67 | 0.505 |
| Age | 0.150 | 0.100 | 1.51 | 0.131 | -0.046 | 0.094 | -0.49 | 0.626 |
| Driver's license | -0.605 | 0.575 | -1.05 | 0.293 | 0.487 | 0.540 | -0.90 | 0.367 |
| Driving experience | 0.257 | 0.150 | 1.71 | 0.087 | 0.212 | 0.142 | 1.49 | 0.136 |
| Education | -0.152 | 0.532 | -0.29 | 0.775 | -0.071 | 0.496 | -0.14 | 0.887 |
| Practitioner | 1.442 | 0.415 | 3.48 | *** | 0.295 | 0.383 | 0.77 | 0.441 |
| Yielding strategy_AV1 | -1.652 | 0.191 | -8.67 | *** | — | — | — | — |
| Yielding strategy_AV2 | — | — | — | — | -1.380 | 0.175 | -7.89 | *** |
| Understanding of AVs | -0.017 | 0.057 | -0.30 | 0.765 | -0.028 | 0.053 | -0.53 | 0.598 |
| Trust in AVs | 0.118 | 0.043 | 2.77 | ** | 0.062 | 0.040 | 1.55 | 0.121 |
| Risk perception of AVs | 0.011 | 0.047 | 0.24 | 0.812 | 0.034 | 0.045 | 0.75 | 0.456 |
| Understanding of eHMIs | -0.207 | 0.067 | -3.08 | ** | -0.066 | 0.063 | -1.04 | 0.298 |
| Trust in eHMIs | 0.146 | 0.055 | 2.66 | ** | 0.017 | 0.051 | 0.34 | 0.737 |
| Crossing focus | 0.091 | 0.056 | 1.62 | 0.105 | 0.042 | 0.053 | 0.81 | 0.419 |
| Crossing adherence | 0.044 | 0.047 | 0.94 | 0.346 | 0.052 | 0.045 | 1.15 | 0.252 |
| Attitudes toward others | -0.070 | 0.047 | -1.47 | 0.141 | -0.080 | 0.045 | -1.77 | 0.076 |
| AIC | | | 924.195 | | | | 1026.141 | |
| BIC | | | 1074.500 | | | | 1176.446 | |
| Log likelihood | | | -430.098 | | | | -481.070 | |

* p ≤ 0.05, ** p ≤ 0.01, *** p ≤ 0.001

## 5. Discussion

This study investigated how eHMIs displayed on AVs influence the crossing decisions of pedestrians in complex interaction scenarios. Two types of eHMIs were examined based on the communication perspective: allocentric eHMIs and egocentric eHMIs. Egocentric communication provides advisory information from the pedestrian's perspective, such as "Walk" and "Stop", while allocentric communication offers referential advice from the vehicle's perspective, such as "Driving" and "Braking". The impact of different eHMI combinations on the crossing behavior of pedestrians across various scenarios was evaluated, focusing on the potential conflicts and interference caused by multi-source information in complex interactions, in addition to comparing the two eHMI types. Additionally, the relationship between subjective cognitive risk and objective behavioral risk when pedestrians interacted with different eHMI combinations were explored, gaining insights that can guide the design and application of future eHMIs.

### 5.1 Cognitive loads on pedestrians due to eHMIs

First, the following question was addressed: Do combinations of vehicles featuring different eHMI configurations increase the cognitive load on pedestrians during crossing?

Considering no eHMIs on vehicles in either lane as the reference condition, the analysis of objective behavioral indicators reveals that the presence of an eHMI on vehicles in a single lane significantly reduced the pedestrian gaze time for the Lane 1 vehicles, as well as the pre-crossing wait time. This suggests that in scenarios involving single-source signals, pedestrians experience lower cognitive load. These observations corroborate the findings of Mührmann et al. (2019), who demonstrated the potential benefits of eHMIs under appropriate conditions, in addition to validating earlier research suggesting that eHMIs can improve the decision-making efficiency of pedestrians.

Focusing more specifically on the active vehicle interacting with the pedestrian, egocentric eHMIs were more effective than allocentric eHMIs in reducing both the gaze time and wait time. Additionally, the analysis of the gaze proportion for the eHMI of the Lane 1 vehicle revealed that compared with the baseline condition (allocentric eHMI on Lane 1 AV and no eHMI on Lane 2 AV), egocentric eHMIs elicited significantly lower gaze proportions; this suggests that pedestrians prefer and more efficiently process the direct behavioral guidance offered by egocentric eHMIs compared with the motion-based cues provided by allocentric eHMIs.

Regarding the interaction effects of dual-lane eHMI configurations, the gaze time and wait time both significantly increased when eHMIs were present on vehicles in both lanes. This indicates that multi-source information may introduce redundancy and raise the cognitive cost of integrating information (Tran et al., 2024), thereby increasing cognitive load significantly.

The findings related to subjective cognitive load, measured using the NASA-TLX scale, mirrored the patterns observed based on the behavioral indicators. Single-lane eHMI configurations were associated with a lower perceived workload, whereas the interaction effects of dual-lane eHMIs resulted in significantly higher subjective cognitive loads. Furthermore, the analysis of the Van Der Laan acceptance scale scores reinforced the conclusion that egocentric eHMIs are more suitable than allocentric ones, particularly in terms of clarity and usability.

The analysis of individual differences captured through the post-experiment questionnaire further revealed that participants more familiar with eHMIs and exhibiting more compliant crossing behavior reported significantly lower cognitive loads. This could be because individuals more knowledgeable about eHMIs tend to trust the information more readily and require less visual confirmation, while rule-abiding pedestrians may rely more on established heuristics and structured decision rules, which reduces the need for cognitive conflict resolution. Additionally, yielding vehicles significantly reduced cognitive load as they provided pedestrians with clear behavioral cues, thus minimizing the need for excessive attention.

Importantly, the effect of age on the cognitive load indicators showed divergent patterns: older pedestrians exhibited shorter gaze time for the Lane 1 vehicles but longer pre-crossing wait times. This suggests that while older individuals may process visual information more efficiently and make faster perceptual judgments, they tend to adopt more cautious and conservative behavioral strategies. These findings highlight the distinction between perceptual processing and behavioral decision-making, reinforcing the idea that cognitive load is a multidimensional construct reflected in diverse behavioral mechanisms and should not be interpreted as a singular phenomenon.

**5.2 Distraction effect**

The second question is: Does information from vehicles in other lanes distract pedestrians? As noted by Lee and Daimon (2025), prolonged exposure to eHMI information may lead to a redistribution of attention for pedestrians, causing them to neglect the surrounding traffic environment and thereby increasing the risk of accidents. We assessed pedestrian distraction using two objective behavioral indicators: the proportion of gaze directed at the vehicle in Lane 2 before crossing and the proportion of gaze directed at the eHMI on the Lane 2 vehicle.

First, considering vehicles in both lanes having no eHMI as the reference group, it was found that the presence of an eHMI on a Lane 2 vehicle significantly increased the gaze proportion for that eHMI. Specifically, the increase was more pronounced for allocentric eHMIs than for egocentric eHMIs. Further analysis considering vehicles in Lane 1 having no eHMI and those in Lane 2 having allocentric eHMIs as the baseline showed that egocentric eHMIs on Lane 2 vehicles significantly reduced the gaze proportion for the eHMI; this reaffirms the findings of Eisma et al. (2019), who reported that allocentric eHMIs are more likely to induce distraction than egocentric eHMIs.

Regarding the interaction effects of different eHMI configurations across the two lanes, relative to the reference condition (no eHMI on vehicles in either lane), configurations where neither lane had vehicles displaying allocentric eHMIs significantly increased the gaze proportion for the vehicle in Lane 2. This suggests that allocentric eHMIs on Lane 2 vehicles significantly contributed to pedestrian distraction. Additionally, considering vehicles with no eHMI in Lane 1 and vehicles with allocentric eHMIs in Lane 2 as the reference scenario, the dual-lane egocentric eHMI configuration significantly reduced the gaze proportion for the Lane 2 vehicle's eHMI. This again supports the conclusion that allocentric eHMIs are more distracting than egocentric ones.

The analysis of individual differences based on the post-experiment questionnaire responses revealed that distraction levels were significantly higher among pedestrians who were older, had more driving experience, or expressed greater trust in AVs. This could be because older pedestrians and those with more driving experience tend to monitor the entire road environment more broadly during crossing. In contrast, those with greater trust in AVs may feel more confident in the actions of the currently interacting vehicle, which allows them to shift part of their attention to vehicles in non-interacting lanes.

In complex traffic environments, attentional shifts and distraction triggered by conflicting information may elevate the risk of collisions. Therefore, future eHMI designs should place greater emphasis on the consistency of the information displayed and should employ appropriate modality configurations to reduce distraction and enhance pedestrian safety.

**5.3 Potential misleading effect**

The third research question addressed in this study was as follows: Would multiple eHMI-equipped AVs potentially mislead pedestrians during street crossings in multi-lane environments?

Lau et al. (2021) found that dynamic eHMIs conveying ambiguous yielding intentions and contradictory information significantly increase the crossing risk for pedestrians. Building upon this, two types of misleading behaviors were examined: Misleading Behavior 1—misleading triggered by yielding information from a Lane 1 vehicle; Misleading Behavior 2—misleading triggered by yielding information from a Lane 2 vehicle.

First, significant differences in how various types of eHMIs affect misleading behavior were identified. Under Misleading Behavior 1, the eHMI type on the Lane 1 AV significantly influenced the likelihood of pedestrians being misled. Considering allocentric eHMIs on Lane 1 AVs as the reference condition, the presence of egocentric eHMIs significantly increased the occurrence of misleading behavior. This suggests that egocentric eHMIs, being easier for pedestrians to interpret, carry a higher risk of inducing misjudgments than allocentric eHMIs. A similar pattern was found under Misleading Behavior 2, where the eHMI type on the Lane 2 AV had a significant main effect on the occurrence of misleading behavior; egocentric eHMIs were again associated with a significantly higher likelihood of misleading pedestrians than allocentric eHMIs.

Second, it was found that the presence of eHMI information on the currently interacting vehicle's lane (i.e., the lane from which the misleading message is not sent) could help mitigate the occurrence of misleading behavior. Under Misleading Behavior 1, the interaction effect between the eHMI configurations of the vehicles in the two lanes significantly reduced the incidence of misleading. Moreover, when the vehicles in both lanes had the same type of eHMI, egocentric eHMIs were more effective than allocentric eHMIs in reducing misleading behavior. A reasonable explanation for this observation is that when the vehicle in the interacting lane also presents eHMI information, pedestrians are less likely to over-rely on the eHMI of the vehicle in the non-interacting lane and shift their decision-making strategy. Furthermore, egocentric eHMIs are generally easier for pedestrians to interpret, which may alleviate confusion and suppress misleading behavior further. This inference is consistent with the findings of Eisele et al. (2024), who emphasized that overreliance on eHMIs can result in misleading effects and potentially adverse outcomes.

Additionally, individual differences significantly influenced the risk of pedestrians being misled. Older pedestrians were more susceptible to misleading, whereas male pedestrians, those who were more focused during crossing, and those with greater knowledge of AVs demonstrated stronger resistance to such effects. In summary, when designing and deploying eHMI systems, signal consistency and cross-lane coordination mechanisms must be carefully considered to prevent the risks of misleading behavior stemming from information imbalance or incomplete communication.

**5.4 Risks of AV–pedestrian interactions**

The final research question to be addressed was this: In multi-vehicle interaction environments, do combinations of different eHMIs significantly increase the crossing risk for pedestrians?

Subramanian et al. (2024) reported that eHMIs may help pedestrians initiate crossing faster, thereby improving crossing efficiency. However, if the judgment is incorrect, the gain in efficiency may be accompanied by an increased risk. In this regard, the findings across the three dimensions were integrated—cognitive load, distraction effects, and misleading behavior—into the modeling of binary high-risk crossing events, defined based on PET thresholds, to comprehensively assess the effects of complex eHMI configurations on pedestrian safety.

Regarding conflicts in Lane 1, the main effect of the eHMI type on Lane 1 significantly increased the crossing risk for pedestrians. Additionally, an allocentric eHMI on a Lane 2 vehicle showed a significant positive main effect.

This indicates that a single eHMI on the vehicle in the interacting lane can substantially elevate the crossing risk and that eHMIs on vehicles in the non-interacting lane may also contribute to higher risks.

Regarding the interaction effects of eHMIs on vehicles in both lanes, two combinations involving allocentric eHMIs on Lane 2 vehicles showed significant negative effects. This suggests that while eHMIs on AVs in the interacting lane increase risk, they can simultaneously mitigate the risk posed by eHMI signals on AVs in the non-interacting lane. In complex multi-lane environments, when the vehicle in the interacting lane does not offer clear eHMI guidance, pedestrians may shift their attention toward other lanes to supplement their decision-making, which increases the crossing risk. However, when the AV in the interacting lane is equipped with an eHMI, pedestrians tend to rely on it as their primary decision-making reference, which reduces the dependence on vehicles in non-interacting lanes and significantly lowers the likelihood of conflicts.

In conjunction with the earlier findings, the analysis of cognitive load revealed that longer gaze time for Lane 1 vehicles before crossing were associated with lower crossing risks; this suggests that while increased cognitive load may reduce efficiency, it may also enhance safety to some extent. By contrast, in terms of distraction, a higher proportion of the gaze being directed toward Lane 2 vehicles significantly increased the crossing risk in Lane 1. This result highlights that shifting attention away from the vehicle in the interacting lane is a key risk factor. Furthermore, Misleading Scenario 2—in which pedestrians were influenced by misleading yielding signals from Lane 2 vehicles while interacting with vehicles in Lane 1—significantly increased the risk in Lane 1. This supports the misleading effect hypothesis further, i.e., pedestrians misled by yielding cues from Lane 2 AVs may ignore the risks from Lane 1 AVs, which increases the potential for unsafe interactions.

Regarding conflicts in Lane 2, only the eHMI type on the Lane 2 AVs had a significant effect on pedestrian risk; this again indicates that a single-lane eHMI configuration can increase the crossing risk.

Regarding cognitive load, the indicators related to the gaze of the pedestrians toward vehicles before crossing did not significantly affect the risk in Lane 2. However, the proportion of the gaze directed toward the eHMIs on Lane 2 AVs significantly reduced the crossing risk. This could be explained by the fact that distraction before crossing increased the attention of the pedestrians to Lane 2 vehicles, thereby reducing the risk during this phase of crossing. Furthermore, Misleading Scenario 1—in which pedestrians were influenced by signals from Lane 1 AVs while crossing Lane 2—significantly increased the crossing risk. This observation also echoes the earlier findings, suggesting that pedestrians misled by yielding messages from Lane 1 vehicles may reduce their vigilance toward Lane 2 vehicles, which elevates the potential for unsafe interactions.

**5.5 Limitations and future research scope**

Although this study provides valuable insights into the effects of eHMI-equipped AVs on pedestrian crossing decisions in complex interaction scenarios, several limitations remain.

First, regarding participant composition, most of the individuals were young university students; thus, other pedestrian groups, such as minors or the elderly, may not have been accurately represented. Future research should strive to include individuals from a broader range of age groups.

Second, the study focused exclusively on the decision-making and behavioral patterns of single pedestrians during crossings, disregarding the effects of multiple pedestrians crossing together. As noted by Hübner et al. (2024), the presence of additional pedestrians has a significant impact on behavior and risk perception. To comprehensively evaluate the influence of eHMIs on pedestrian behavior in complex traffic environments, future studies should adopt experimental designs involving multiple pedestrians and vehicles.

Third, the experiment did not consider vehicle-related factors, such as vehicle type (e.g., passenger car vs. truck) or vehicle speed, both of which could significantly affect the risk perception and behavior of pedestrians (Ye et al., 2024). In real-world traffic, perceived threat or safety levels can vary across vehicle types, while vehicle speed is a well-known determinant of crossing decisions. Future research should incorporate a wider range of vehicle characteristics to improve the ecological validity of the findings.

Finally, the experiment was conducted in a VR simulation environment. This advanced platform not only enabled immersive interaction but also generated rich, high-resolution behavioral data that would be difficult to capture in real-world settings. However, VR cannot fully reproduce the complexities and unpredictability of real traffic. The participants also performed multiple trials in similar environments, which may have rendered the virtual traffic situations more predictable than real-world conditions. Future research should aim to introduce more varied scene parameters to diversify the traffic scenarios and reduce the potential biases introduced by the virtual environment.

## 6. Conclusions

This study explored the mechanisms through which different types and combinations of eHMIs on AVs affect the crossing behavior of pedestrians in complex multi-lane interaction scenarios. By examining four key parameters, namely cognitive load, distraction effects, misleading behavior, and crossing risk, the potential interference caused by eHMIs during AV–pedestrian interactions was determined.

The findings indicate that in single-lane scenarios, eHMIs can effectively reduce the cognitive load on pedestrians and simplify their crossing decisions. However, in dual-lane environments, the presence of eHMIs significantly increased both cognitive load and distraction, particularly in the case of lanes equipped with different perspectives of eHMIs or inconsistent signal types. Under these conditions, pedestrians were more easily distracted. Furthermore, allocentric eHMIs induced higher cognitive loads and greater distraction than egocentric eHMIs, which suggests that pedestrians are more inclined to respond to behavioral guidance rather than interpret the intentions of vehicles.

Regarding misleading behavior, the cross-lane misinterpretation mechanisms resulting from different eHMI combinations were quantified. It was demonstrated that pedestrians were misled primarily by yielding signals from eHMIs displayed on AVS in non-interacting lanes, especially under asymmetric signal conditions. When no eHMI was present on the interacting vehicle, pedestrians tended to rely on eHMI signals from adjacent, non-interacting vehicles and mistakenly transferred that decision strategy to the actual interacting lane. This often led to a misjudgment of the traffic environment, thus increasing the risk of unsafe interactions. Further analysis revealed that under such conditions, egocentric eHMIs were more likely to induce misjudgments than allocentric eHMIs, which highlights the importance of maintaining signal symmetry and consistency in the deployment of eHMIs across multi-lane scenarios.

Finally, by integrating the findings related to cognitive load, distraction, and misleading behavior into the analysis of crossing risk, it was found that in complex multi-lane environments, increased cognitive demands, attentional shifts, and decision-making errors induced by eHMIs significantly elevate the likelihood of high-risk vehicle–pedestrian interactions. These findings underscore the necessity of ensuring consistency in eHMI types and coordination across lanes to mitigate the risks of potential collisions during pedestrian crossings.

While this study provides empirical evidence of eHMI-related interference and risk in multi-vehicle interaction settings, there are certain limitations. Additionally, the results emphasize the role of individual differences in pedestrian responses. Future research should therefore broaden the participant sample, incorporate multi-pedestrian and multi-vehicle scenarios, and introduce more contextual variables to ensure a more comprehensive assessment of the impact of eHMIs on pedestrian behavior.

# References

Asaithambi, G., Kuttan, M. O., & Chandra, S. (2016). Pedestrian road crossing behavior under mixed traffic conditions: A comparative study of an intersection before and after implementing control measures. *Transportation in Developing Economies*, 2(2), 14.

Barendse, M. (2019). External human-machine interfaces on autonomous vehicles: The effects of information type on pedestrian crossing decisions. *Delft University of Technology*.

Berchialla, P., Baldi, I., Notaro, V., Barone-Monfrin, S., Bassi, F., & Gregori, D. (2009). Flexibility of Bayesian generalized linear mixed models for oral health research. *Statistics in Medicine*, 28(28), 3509–3522.

Betz, J., Lutwitzi, M., & Peters, S. (2024). A new taxonomy for automated driving: structuring applications based on their operational design domain, level of automation and automation readiness. *arXiv preprint arXiv:2404.17044*.

Bouchard, S., Robillard, G., & Renaud, P. (2007). Revising the factor structure of the simulator sickness questionnaire. *Annual Review of Cybertherapy and Telemedicine*, 5, 128–137.

Bungum, T. J., Day, C., & Henry, L. J. (2005). The association of distraction and caution displayed by pedestrians at a lighted crosswalk. *Journal of Community Health*, 30(4), 269–279.

Campisi, T., Otković, I. I., Šurdonja, S., & Deluka-Tibljaš, A. (2022). Impact of social and technological distraction on pedestrian crossing behaviour: A case study in Enna, Sicily. *Transportation Research Procedia*, 60, 100–107.

Carmona, J., Guindel, C., Garcia, F., & De La Escalera, A. (2021). eHMI: Review and guidelines for deployment on autonomous vehicles. *Sensors*, 21(9), 2912.

Chang, C. M., Toda, K., Igarashi, T., Miyata, M., & Kobayashi, Y. (2018, September). A video-based study comparing communication modalities between an autonomous car and a pedestrian. In *Adjunct Proceedings of the 10th International Conference on Automotive User Interfaces and Interactive Vehicular Applications* (pp. 104–109).

Chang, C. M., Toda, K., Sakamoto, D., & Igarashi, T. (2017, September). Eyes on a car: An interface design for communication between an autonomous car and a pedestrian. In *Proceedings of the 9th International Conference on Automotive User Interfaces and Interactive Vehicular Applications* (pp. 65–73).

Chen, X., Li, X., Hou, Y., Yang, W., Dong, C., & Wang, H. (2025). Effect of eHMI-equipped automated vehicles on pedestrian crossing behavior and safety: A focus on blind spot scenarios. *Accident Analysis & Prevention*, 212, 107915.

Clamann, M., Aubert, M., & Cummings, M. L. (2017, January). Evaluation of vehicle-to-pedestrian communication displays for autonomous vehicles. *The 96th Annual Meeting of Transportation Research Board*, Washington DC, United States.

Colley, M., Bajrovic, E., & Rukzio, E. (2022, April). Effects of pedestrian behavior, time pressure, and repeated exposure on crossing decisions in front of automated vehicles equipped with external communication. In *Proceedings of the 2022 CHI Conference on Human Factors in Computing Systems* (pp. 1–11).

Cooper, P. J. (1984). Experience with traffic conflicts in Canada with emphasis on "post encroachment time" techniques. In E. Asmussen (Ed.), *International Calibration Study of Traffic Conflict Techniques* (pp. 75–96). Springer Berlin Heidelberg.

Das, P. (1998). *Econometrics in theory and practice*. Springer.

Deb, S., Strawderman, L., Carruth, D. W., DuBien, J., Smith, B., & Garrison, T. M. (2017). Development and validation of a questionnaire to assess pedestrian receptivity toward fully autonomous vehicles. *Transportation Research Part C: Emerging Technologies*, 84, 178–195.

de Winter, J., & Dodou, D. (2022). External human–machine interfaces: Gimmick or necessity? *Transportation Research Interdisciplinary Perspectives*, 15, 100643.

Dey, D., Matviienko, A., Berger, M., Pfleging, B., Martens, M., & Terken, J. (2021). Communicating the intention of an automated vehicle to pedestrians: The contributions of eHMI and vehicle behavior. *it - Information Technology*, 63(2), 123–141.

Dommes, A. (2019). Street-crossing workload in young and older pedestrians. *Accident Analysis & Prevention*, 128, 175–184.

Eisele, D., Kraus, J., Schlemer, M. M., & Petzoldt, T. (2024). Should automated vehicles communicate their state or intent? Effects of eHMI activations and non-activations on pedestrians' trust formation and crossing behavior. *Multimedia Tools and Applications*, Advance online publication.

Eisma, Y. B., Reiff, A., Kooijman, L., Dodou, D., & de Winter, J. C. F. (2021). External human-machine interfaces: Effects of message perspective. *Transportation Research Part F: Traffic Psychology and Behaviour*, 78, 30–41.

Eisma, Y. B., van Bergen, S., Ter Brake, S. M., Hensen, M. T. T., Tempelaar, W. J., & de Winter, J. C. (2019). External human–machine interfaces: The effect of display location on crossing intentions and eye movements. *Information*, 11(1), 13.

Esmaili, A., Aghabayk, K., Parishad, N., & Stephens, A. N. (2021). Investigating the interaction between pedestrian behaviors and crashes through validation of a pedestrian behavior questionnaire (PBQ). *Accident Analysis & Prevention*, 153, 106050.

Febiyani, A., Febriani, A., & Ma'Sum, J. (2021). Calculation of mental load from e-learning student with NASA TLX and SOFI method. *Jurnal Sistem Dan Manajemen Industri*, 5(1), 35–42.

Ferenchak, N. N., & Shafique, S. (2022). Pedestrians' perceptions of autonomous vehicle external human-machine interfaces. *ASCE-ASME Journal of Risk and Uncertainty in Engineering Systems, Part B: Mechanical Engineering*, 8(3), 034501.

Field, A. (2024). *Discovering Statistics Using IBM SPSS Statistics*. SAGE Publications.

Geruschat D. R., Hassan S. E., & Turano K. A. (2003). Gaze behavior while crossing complex intersections. *Optometry and Vision Science*, 80(7), 515.

Gruden, C., Ištoka Otković, I., & Šraml, M. (2021). Pedestrian safety at roundabouts: Their crossing and glance behavior in the interaction with vehicular traffic. *Accident Analysis & Prevention*, 159, 106290.

Guo, J., Yuan, Q., Yu, J., Chen, X., Yu, W., Cheng, Q., Wang, W., Luo, W., & Jiang, X. (2022). External human–machine interfaces for autonomous vehicles from pedestrians' perspective: A survey study. *Sensors*, 22(9), 3339.

Habibovic, A., Lundgren, V. M., Andersson, J., Klingegård, M., Lagström, T., Sirkka, A., ..., & Larsson, P. (2018). Communicating intent of automated vehicles to pedestrians. *Frontiers in Psychology*, 9, 1336.

Holländer, K., Wintersberger, P., & Butz, A. (2019, September). Overtrust in external cues of automated vehicles: an experimental investigation. In *Proceedings of the 11th International Conference on Automotive User Interfaces and Interactive Vehicular Applications* (pp. 211–221).

Hossain, M. M., Zhou, H., Sun, X., Hossain, A., & Das, S. (2024). Crashes involving distracted pedestrians: Identifying risk factors and their relationships to pedestrian severity levels and distraction modes. *Accident Analysis & Prevention*, 194, 107359.

Howlader, M. M., Mannering, F., & Haque, M. M. (2024). Estimating crash risk and injury severity considering multiple traffic conflict and crash types: A bivariate extreme value approach. *Analytic Methods in Accident Research*, 42, 100331.

Hsiao, C. (2007). Panel data analysis—Advantages and challenges. *Test*, 16(1),

1–22.

Islam, Z., Abdel-Aty, M., Goswamy, A., Abdelraouf, A., & Zheng, O. (2023). Effect of signal timing on vehicles' near misses at intersections. *Scientific Reports*, 13(1), 9065.

Izquierdo, R., Martín, S., Alonso, J., Parra, I., Sotelo, M. A., & Fernaádez–Llorca, D. (2023, September). Human-vehicle interaction for autonomous vehicles in crosswalk scenarios: Field experiments with pedestrians and passengers. In *2023 IEEE 26th International Conference on Intelligent Transportation Systems (ITSC)* (pp. 2473–2478). IEEE.

Kaleefathullah, A. A., Merat, N., Lee, Y. M., Eisma, Y. B., Madigan, R., Garcia, J., & Winter, J. de. (2022). External human–machine interfaces can be misleading: An examination of trust development and misuse in a CAVE-based pedestrian simulation environment. *Human Factors*, 64(6), 1070–1085.

Kellogg, R. S., Kennedy, R. S., & Graybiel, A. (1965). Motion sickness symptomatology of labyrinthine defective and normal subjects during zero gravity maneuvers. *Aerospace Medicine*, 36, 315–318.

Kooijman, L., Happee, R., & de Winter, J. C. F. (2019). How do eHMIs affect pedestrians' crossing behavior? A study using a head-mounted display combined with a motion suit. *Information*, 10(12), Article 12.

Krishna, K. V., Kapruwan, R., & Choudhary, P. (2024). Understanding distracted pedestrians' risky behaviour: The role of walking and visual characteristics through a field study. *Transportation Research Part F: Traffic Psychology and Behaviour*, 101, 111–129.

Lau, M., Jipp, M., & Oehl, M. (2021, September). Investigating the interplay between eHMI and dHMI for automated buses: How do contradictory signals influence a pedestrian's willingness to cross?. In *13th International Conference on Automotive User Interfaces and Interactive Vehicular Applications* (pp. 152–155).

Lee, J., & Daimon, T. (2025). Not always good: Mitigating pedestrians' less careful crossing behavior by external human-machine interfaces on automated vehicles. *International Journal of Human–Computer Interaction*, 41(8), 4528–4540.

Lee, Y. M., Sidorov, V., Madigan, R., Garcia de Pedro, J., Markkula, G., & Merat, N. (2025). Hello, is it me you're stopping for? The effect of external human machine interface familiarity on pedestrians' crossing behaviour in an ambiguous situation. *Human Factors*, 67(3), 264–279.

Li, X., You, Z., Ma, X., Pang, X., Min, X., & Cui, H. (2024). Effect of autonomous vehicles on car-following behavior of human drivers: Analysis based on structural equation models. *Physica A: Statistical Mechanics and Its Applications*, 633, 129360.

Liu, H., & Hirayama, T. (2025). Pre-instruction for pedestrians interacting autonomous vehicles with eHMI: Effects on their psychology and walking behavior. *IEEE Transactions on Intelligent Transportation Systems*. Advance online publication.

Liu, H., Hirayama, T., Morales Saiki, L. Y., & Murase, H. (2023). Implicit interaction with an autonomous personal mobility vehicle: Relations of pedestrians' gaze behavior with situation awareness and perceived risks. *International Journal of Human–Computer Interaction*, 39(10), 2016–2032.

Liu, H., Hirayama, T., Saiki, L. Y. M., & Murase, H. (2020, September). What timing for an automated vehicle to make pedestrians understand its driving intentions for improving their perception of safety?. In *2020 IEEE 23rd International Conference on Intelligent Transportation Systems (ITSC)* (pp. 1–6). IEEE.

Lyu, W., Zhang, W., Wang, X., Ding, Y., & Yang, X. (2024). Pedestrians' responses to scalable automated vehicles with different external human-machine interfaces: Evidence from a video-based eye-tracking experiment. *Transportation Research Part F: Traffic Psychology and Behaviour*, 103, 112–127.

Mohammed, H. A. (2021). Assessment of distracted pedestrian crossing behavior at midblock crosswalks. *IATSS Research*, 45(4), 584–593.

Mührmann, L. (2019). *Exploration of the communication between autonomous vehicles (AVs) and pedestrians via exterior human-machine interfaces (eHMIs)* [Master's Thesis, University of Twente].

O'Dell, A., Morris, A., Filtness, A., & Barnes, J. (2023). The impact of pedestrian distraction on safety behaviours at controlled and uncontrolled crossings. *Future Transportation*, 3(4), 1195–1208.

Raoniar, R., & Maurya, A. K. (2024). Digital and social distractions' impact on pedestrian road crossing behavior at signalized intersection crosswalks. *Transportation Letters*, 16(7), 672–687.

Rasouli, A., & Tsotsos, J. K. (2019). Autonomous vehicles that interact with pedestrians: A survey of theory and practice. *IEEE Transactions on Intelligent Transportation Systems*, 21(3), 900–918.

Rosenholtz, R., Li, Y., & Nakano, L. (2007). Measuring visual clutter. *Journal of Vision*, 7(2), 17.

Sarstedt, M., Ringle, C. M., & Hair, J. F. (2022). Partial least squares structural equation modeling. In *Handbook of Market Research* (pp. 587–632). Springer, Cham.

Schneider, S., Ratter, M., & Bengler, K. (2019, October). *Pedestrian behavior in virtual reality: Effects of gamification and distraction* [Conference presentation]. Road Safety and Simulation International Conference, Iowa City, IA, United States.

Schwebel, D. C., Canter, M. F., Hasan, R., Griffin, R., White, T. R., & Johnston, A. (2022). Distracted pedestrian behavior: An observational study of risk by situational environments. *Traffic Injury Prevention*, 23(6), 346–351.

Song, Y., Jiang, Q., Chen, W., Zhuang, X., & Ma, G. (2023). Pedestrians' road-crossing behavior towards eHMI-equipped autonomous vehicles driving in segregated and mixed traffic conditions. *Accident Analysis & Prevention*, 188, 107115.

Subramanian, L. D., O'Neal, E. E., Kim, N. Y., Noonan, M., Plumert, J. M., & Kearney, J. K. (2024). Deciding when to cross in front of an autonomous vehicle: how child and adult pedestrians respond to eHMI timing and vehicle kinematics. *Accident Analysis & Prevention*, 202, 107567.

Sweller, J. (2011). Cognitive load theory. In *Psychology of Learning and Motivation* (Vol. 55, pp. 37–76). Elsevier.

Tapiro, H., Oron-Gilad, T., & Parmet, Y. (2020). Pedestrian distraction: The effects of road environment complexity and age on pedestrian's visual attention and crossing behavior. *Journal of Safety Research*, 72, 101–109.

Thompson, L. L., Rivara, F. P., Ayyagari, R. C., & Ebel, B. E. (2013). Impact of social and technological distraction on pedestrian crossing behaviour: An observational study. *Injury Prevention*, 19(4), 232–237.

Tian, K., Markkula, G., Wei, C., Sadraei, E., Hirose, T., Merat, N., & Romano, R. (2022). Impacts of visual and cognitive distractions and time pressure on pedestrian crossing behaviour: A simulator study. *Accident Analysis & Prevention*, 174, 106770.

Tran, T. T. M., Parker, C., & Tomitsch, M. (2021). A review of virtual reality studies on autonomous vehicle–pedestrian interaction. *IEEE Transactions on Human-Machine Systems*, 51(6), 641–652.

Tran, T. T. M., Parker, C., Hoggenmüller, M., Wang, Y., & Tomitsch, M. (2024, May). Exploring the impact of interconnected external interfaces in autonomous vehicles on pedestrian safety and experience. In *Proceedings of the 2024 CHI Conference on Human Factors in Computing Systems* (pp. 1–17).

Van Der Laan, J. D., Heino, A., & De Waard, D. (1997). A simple procedure for the assessment of acceptance of advanced transport telematics. *Transportation Research Part C: Emerging Technologies*, 5(1), 1–10.

Verkuilen, J., & Smithson, M. (2012). Mixed and mixture regression models for continuous bounded responses using the beta distribution. *Journal of Educational and Behavioral Statistics*, 37(1), 82–113.

von Janczewski, N., Kraus, J., Engeln, A., & Baumann, M. (2022). A subjective one-item measure based on NASA-TLX to assess cognitive workload in driver-vehicle interaction. *Transportation Research Part F: Traffic Psychology and Behaviour*, 86, 210–225.

Wang, H., Li, D., Wang, Q., Schwebel, D. C., Miao, L., & Shen, Y. (2022). How distraction affects pedestrian response: Evidence from behavior patterns and cortex oxyhemoglobin changes. *Transportation Research Part F: Traffic Psychology and Behaviour*, 91, 414–430.

Wang, K., Li, G., Chen, J., Long, Y., Chen, T., Chen, L., & Xia, Q. (2020). The adaptability and challenges of autonomous vehicles to pedestrians in urban China. *Accident Analysis & Prevention*, 145, 105692.

Weiß, S. L., Eisele, D., & Petzoldt, T. (2022). External human-machine-interfaces on automated vehicles: Which message and perspective do pedestrians in crossing situations understand best? In *Intelligent human systems integration (IHSI 2022): Integrating people and intelligent systems (Vol. 22)*. AHFE International.

Wilbrink, M., Lau, M., Illgner, J., Schieben, A., & Oehl, M. (2021). Impact of external human–machine interface communication strategies of automated vehicles on pedestrians' crossing decisions and behaviors in an urban environment. *Sustainability*, 13(15), 8396.

Wilbrink, M., Nuttelmann, M., & Oehl, M. (2021, September). Scaling up automated vehicles' eHMI communication designs to interactions with multiple pedestrians–putting eHMIs to the test. In *13th International Conference on Automotive User Interfaces and Interactive Vehicular Applications* (pp. 119–122).

Williams, L. J., & Abdi, H. (2010). Fisher's least significant difference (LSD) test. *Encyclopedia of research design*, *218*(4), 840-853.

World Health Organization. 2023. *Global status report on road safety 2023*. World Health Organization. https://www.who.int/southeastasia/publications/i/item/9789240086517

Yang, J., Rahardja, S., & Fränti, P. (2019, December). Outlier detection: How to threshold outlier scores?. In *Proceedings of the International Conference on Artificial Intelligence, Information Processing and Cloud Computing* (pp. 1–6).

Ye, Y., Che, Y., & Liang, H. (2024). Exploring the influence of pedestrian attitude, propensity, and risk perception on gap acceptance between platooning autonomous trucks. In *2024 IEEE 27th International Conference on Intelligent Transportation Systems (ITSC)* (pp. 3645–3650).

Ye, Y., Wong, S. C., Li, Y. C., & Choi, K. M. (2023). Crossing behaviors of drunk pedestrians unfamiliar with local traffic rules. *Safety Science*, 157, 105924.

Ye, Y., Wong, S. C., Li, Y. C., & Lau, Y. K. (2020). Risks to pedestrians in traffic systems with unfamiliar driving rules: A virtual reality approach. *Accident Analysis & Prevention*, 142, 105565.
76

Zheng, L., Ismail, K., & Meng, X. (2016). Investigating the heterogeneity of postencroachment time thresholds determined by peak over threshold approach. *Transportation Research Record*, 2601(1), 17–23.

Zhu, H., Han, T., Alhajyaseen, W. K., Iryo-Asano, M., & Nakamura, H. (2022). Can automated driving prevent crashes with distracted pedestrians? An exploration of motion planning at unsignalized mid-block crosswalks. *Accident Analysis & Prevention*, 173, 106711.

## Appendix A

Table A1. Outline of NASA-TLX questionnaire.

| Indicator | Description |
|---|---|
| Mental Demand (MD) | How much mental and perceptual activity was required (e.g., thinking, deciding, calculating, remembering, looking, or searching)? Was the task easy or demanding, simple or complex, exacting or forgiving? |
| Physical Demand (PD) | How much physical activity was required (e.g., pushing, pulling, turning, controlling, or activating)? Was the task easy or demanding, slow or brisk, slack or strenuous, restful or laborious? |
| Temporal Demand (TD) | How much time pressure did you feel due to the rate or pace at which the tasks or task elements appeared? Was the pace slow and leisurely or rapid and frantic? |
| Own Performance (OP) | How successful do you think you were in accomplishing the goals of the task set by the experimenter (or yourself)? How satisfied were you with your performance in accomplishing these goals? |
| Effort (EF) | How hard did you have to work (mentally and physically) to accomplish your level of performance? |
| Frustration Level (FR) | How insecure, discouraged, irritated, stressed, and annoyed versus secure, gratified, content, relaxed, and complacent did you feel during the task? |

Table A2. Interpretation scores for NASA-TLX.

| Workload | Value |
|---|---|
| Low | 0–9 |
| Medium | 10–29 |
| Somewhat high | 30–49 |
| High | 50–79 |
| Very high | 80–100 |

Table A3. Weighted scores for NASA-TLX.

| MD☐ | PD☐ | TD☐ |
|---|---|---|

| PD☐ | TD☐ | EF☐ |
| MD☐ | PD☐ | TD☐ |
| TD☐ | OP☐ | FR☐ |
| MD☐ | PD☐ | OP☐ |
| OP☐ | EF☐ | EF☐ |
| MD☐ | PD☐ | OP☐ |
| EF☐ | FR☐ | FR☐ |
| MD☐ | TD☐ | EF☐ |
| FR☐ | OP☐ | FR☐ |

**Table A4.** Van Der Laan scale.

Allocentric communication: Primarily conveys the current driving status of the vehicle, such as "Driving" or "Braking".

I think this type of eHMI communication is...

| | | | | |
|---|---|---|---|---|
| useless | | | | useful |
| bad | | | | good |
| superfluous | | | | effective |
| worthless | | | | assisting |
| sleep-inducing | | | | alertness-raising |
| unpleasant | | | | pleasant |
| annoying | | | | nice |
| irritating | | | | likeable |
| undesirable | | | | desirable |

Egocentric communication: Primarily conveys the desired pedestrian behavior from the vehicle's perspective, such as "Stop" or "Walk".

I think this type of eHMI communication is...

| | | | | |
|---|---|---|---|---|
| useless | | | | useful |
| bad | | | | good |
| superfluous | | | | effective |
| worthless | | | | assisting |
| sleep-inducing | | | | alertness-raising |
| unpleasant | | | | pleasant |

| annoying | | | | nice |
|---|---|---|---|---|
| irritating | | | | likeable |
| undesirable | | | | desirable |

**Appendix B**

Measurement of risk perception

Survey Questionnaire

A. Demographic Background

1. What is your age? (Fill in the blank)
2. What is your gender? (Male, Female)
3. What is your name?
4. Do you have a driver's license? (Yes, No)
5. How long have you been driving? (No driving experience, Less than 1 year, 1–3 years, 3–5 years, More than 5 years)
6. What is your highest level of education (including current studies)? (Below middle school, High school, Vocational school, Junior college, Bachelor's degree, Master's degree or above)
7. Are you a researcher or practitioner in the transportation field (e.g., logistics, traffic planner, researcher in transportation studies)? (Yes, No)

B. Autonomous Vehicles

(1) Knowledge

Autonomous vehicles (also known as self-driving cars) are intelligent vehicles that operate without human intervention, relying on AI, computer vision, radar, monitoring systems, and GPS to safely drive themselves.

1. Before reading the description above, I already understood the principles of autonomous vehicles.
(Strongly disagree, Disagree, Neutral, Agree, Strongly agree)
2. Before reading the description above, I already understood how autonomous vehicles operate and function.
(Strongly disagree, Disagree, Neutral, Agree, Strongly agree)
3. Before reading the description above, I already understood the current advantages and disadvantages of autonomous vehicles (e.g., safer, more convenient).
(Strongly disagree, Disagree, Neutral, Agree, Strongly agree)
4. Before reading the description above, I already understood the history and future prospects of the autonomous vehicle industry.
(Strongly disagree, Disagree, Neutral, Agree, Strongly agree)

(2) Trust

1. After autonomous vehicles are introduced to the market, I plan to use them for my future travels.
(Strongly disagree, Disagree, Neutral, Agree, Strongly agree)

2. I believe autonomous vehicles can improve road safety and reduce traffic accidents.
(Strongly disagree, Disagree, Neutral, Agree, Strongly agree)

3. While using an autonomous vehicle, I wouldn't feel the need to constantly observe the surroundings and prepare to take control.
(Strongly disagree, Disagree, Neutral, Agree, Strongly agree)

4. I would recommend autonomous vehicles to my family and friends and feel confident about them using it.
(Strongly disagree, Disagree, Neutral, Agree, Strongly agree)

5. Compared to non-autonomous vehicles, I would feel more at ease doing other things (e.g., checking emails on my smartphone, chatting with companions) when crossing the road in front of an autonomous vehicle.
(Strongly disagree, Disagree, Neutral, Agree, Strongly agree)

6. I would encourage my family and friends to feel at ease when crossing in front of autonomous vehicles.
(Strongly disagree, Disagree, Neutral, Agree, Strongly agree)

(3) Risk Perception

1. I am concerned about system or equipment failures in autonomous vehicles.
(Strongly disagree, Disagree, Neutral, Agree, Strongly agree)

2. I am concerned about legal responsibility in the event of an accident involving an autonomous vehicle.
(Strongly disagree, Disagree, Neutral, Agree, Strongly agree)

3. I think autonomous vehicles may interfere with manually driven vehicles during operation.
(Strongly disagree, Disagree, Neutral, Agree, Strongly agree)

4. I think autonomous vehicles may negatively affect pedestrians and non-motorized road users.
(Strongly disagree, Disagree, Neutral, Agree, Strongly agree)

5. I think the risk of accidents is relatively high when autonomous vehicles interact with pedestrians.

(Strongly disagree, Disagree, Neutral, Agree, Strongly agree)

C. eHMI (External Human-Machine Interface)

(1) Knowledge

eHMI refers to the external human-machine interface responsible for establishing physical communication between two entities (i.e., the user and the system). Information (i.e., "feedback") is provided via control panels with light signals, display fields, or buttons, or via software on a user device.

1. Before reading the description above, I already understood the concept and function of eHMI.

(Strongly disagree, Disagree, Neutral, Agree, Strongly agree)

2. Before reading the description above, I already understood the development background of eHMI.

(Strongly disagree, Disagree, Neutral, Agree, Strongly agree)

3. Before reading the description above, I already understood the pros and cons of eHMI.

(Strongly disagree, Disagree, Neutral, Agree, Strongly agree)

(2) Trust

1. I believe eHMI can ensure pedestrian safety.

(Strongly disagree, Disagree, Neutral, Agree, Strongly agree)

2. I think eHMI can enhance road safety and reduce traffic accidents.

(Strongly disagree, Disagree, Neutral, Agree, Strongly agree)

3. I intend to install eHMI on my vehicle when it becomes widely available.

(Strongly disagree, Disagree, Neutral, Agree, Strongly agree)

4. I would recommend eHMI to my friends and family.

(Strongly disagree, Disagree, Neutral, Agree, Strongly agree)

D. Pedestrian Crossing Aggressiveness

(1) Attention

1. I chat on my phone or listen to music while crossing the road.

(Strongly disagree, Disagree, Neutral, Agree, Strongly agree)

2. I get distracted by my thoughts and forget to observe traffic before crossing.

(Strongly disagree, Disagree, Neutral, Agree, Strongly agree)

3. I have crossed several streets and intersections without paying attention to traffic.

(Strongly disagree, Disagree, Neutral, Agree, Strongly agree)

4. I sometimes fail to notice other pedestrians or obstacles while walking.

(Strongly disagree, Disagree, Neutral, Agree, Strongly agree)

(2) Compliance

1. I always walk on the right side to avoid bumping into others.

(Strongly disagree, Disagree, Neutral, Agree, Strongly agree)

2. On streets with a central divider, I cross the first half and wait in the middle before crossing the second half.

(Strongly disagree, Disagree, Neutral, Agree, Strongly agree)

3. If there is an underpass or pedestrian bridge, I choose it instead of crossing the road directly.

(Strongly disagree, Disagree, Neutral, Agree, Strongly agree)

4. Even if the pedestrian signal turns green, I wait and observe the traffic before crossing.

(Strongly disagree, Disagree, Neutral, Agree, Strongly agree)

5. On narrow sidewalks, I walk in a line with others to avoid disturbing other pedestrians.

(Strongly disagree, Disagree, Neutral, Agree, Strongly agree)

6. Even when following others across the road, I always observe the traffic.

(Strongly disagree, Disagree, Neutral, Agree, Strongly agree)

(3) Attitudes Toward Other Road Users

1. I appreciate every driver who stops to let me cross.

(Strongly disagree, Disagree, Neutral, Agree, Strongly agree)

2. I don't get angry or shout at other road users (e.g., pedestrians, drivers, cyclists).

(Strongly disagree, Disagree, Neutral, Agree, Strongly agree)

3. I don't deliberately walk slowly across the road to annoy drivers.

(Strongly disagree, Disagree, Neutral, Agree, Strongly agree)

4. I would never damage a driver's car even if I were angry at them.

(Strongly disagree, Disagree, Neutral, Agree, Strongly agree)

5. If a vehicle is approaching, I wouldn't cross the road because I don't think they would stop for me.

(Strongly disagree, Disagree, Neutral, Agree, Strongly agree)

6. If a vehicle has no cars behind it, I would let it go first even if I have the right of way.

(Strongly disagree, Disagree, Neutral, Agree, Strongly agree)

7. I don't force other pedestrians to give way for me when crossing the road.

(Strongly disagree, Disagree, Neutral, Agree, Strongly agree)